\documentclass{iopart}
\usepackage[utf8]{inputenc}
\usepackage{cmap}
\usepackage[T1]{fontenc}

\usepackage{graphicx}

\expandafter\let\csname equation*\endcsname\relax
\expandafter\let\csname endequation*\endcsname\relax
\usepackage{amsmath}
\usepackage{amsthm, amssymb, color}
\usepackage{sistyle}
\usepackage{calc}
\usepackage{tikz}
\usepackage{cite}
\usepackage{multirow}

\newcommand{\lam}{\lambda}
\newcommand{\sig}{\sigma}

\newcommand{\kla}[1]{\left(#1\right)}
\newcommand{\skal}[1]{\left\langle#1\right\rangle}
\newcommand\wt[1]{\widetilde{#1}}
\newcommand\eu[1]{\mathrm{e}^{#1}}
\newcommand\sign{\operatorname{sign}}

\newcommand{\tabcapskip}{\vspace{.3ex}}

\newcommand\subref[1]{??}

\usepackage[acronym]{glossaries}
\newacronym{kpz}{KPZ}{Kardar--Parisi--Zhang}
\newacronym{ew}{EW}{Edwards--Wilkinson}
\newacronym{rsos}{RSOS}{restricted solid-on-solid model}
\newacronym{bd}{BD}{ballistic deposition model}
\newacronym{ms}{MS}{multi-surface coding}
\newacronym{fdr}{FDR}{fluctuation-dissipation relation}
\newacronym{kk}{KK}{Kim--Kosterlitz}
\newacronym{asep}{ASEP}{asymmetric simple exclusion process}
\newacronym{tasep}{TASEP}{totally asymmetric exclusion process}
\newacronym{mbe}{MBE}{molecular beam epitaxy}
\newacronym{ks}{KS}{Kuramoto--Shivashinsky}
\newacronym{lsi}{LSI}{local scale-invariance}
\newacronym{goe}{GOE}{Gaussian orthogonal ensemble}
\newacronym{gue}{GUE}{Gaussian unitary ensemble}
\newacronym{dlg}{DLG}{driven lattice gases}

\newacronym{rng}{RNG}{random number generator}
\newacronym{lcg}{LCG}{linear congruential generator}
\newacronym{gpu}{GPU}{graphics processing unit}
\newacronym{gpgpu}{GPGPU}{general purpose computing on graphics processing units}
\newacronym{cpu}{CPU}{central processing unit}
\newacronym{simt}{SIMT}{single instruction multiple thread}
\newacronym{ram}{RAM}{computer's main memory}
\newacronym{l1}{L1 cache}{level-1-cache}
\newacronym{l3}{L3 cache}{level-3-cache}
\newacronym{lsb}{LSB}{least significant bit}
\newacronym{cu}{VP}{vector processor}
\newacronym{tdp}{TDP}{thermal design power}
\newacronym{simd}{SIMD}{single instruction multiple data}
\newacronym{mpi}{MPI}{Message Passing Interface}

\newacronym{fcc}{\texttt{fcc}}{face-centered cubic}
\newacronym{sc}{\texttt{sc}}{simple cubic}

\newacronym{cuda}{{CUDA}}{Compute unified device architecture}

\newacronym{nn}{NN}{nearest neighbor}
\newacronym{kmc}{KLMC}{3D kinetic Metropolis lattice Monte Carlo}
\newacronym{bkl}{BKL}{Bortz--Kalos--Lebowitz}
\newacronym{mc}{MC}{Monte Carlo}
\newacronym[shortplural=MCS]{mcs}{MCS}{Monte Carlo step}
\newacronym{prng}{PRNG}{pseudo random number generator}
\newacronym{lcrng}{LCRNG}{linear congruential random number generator}
\newacronym{md}{MD}{molecular dynamics}
\newacronym{pbc}{PBC}{periodic boundary conditions}
\newacronym{fbc}{FBC}{fixed boundary conditions}
\newacronym{wtm}{WTM}{waiting time method}

\newacronym{pl}{PL}{power law}
\newacronym{rhs}{r.h.s.}{right-hand side}
\newacronym{resp}{resp.}{respectively}
\newacronym{snr}{S/N}{signal-to-noise ratio}
\newacronym{fom}{FOM}{figure of merit}
\newacronym{ac}{AC}{autocorrelation}
\newacronym{su}{SU}{super-universality}

\newacronym{dd}{DD}{domain decomposition}
\newacronym{cb}{CB}{checker-board}
\newacronym{sca}{SCA}{stochastic cellular automaton}
\newacronym{rs}{RS}{random-sequential}
\newacronym{dt}{DT}{double tiling}
\newacronym{db}{DB}{dead border}
\newacronym{cdb}{cDB}{coarse dead border}
\newacronym{dtr}{DTr}{\gls{dt} \gls{dd} with random origin}
\newacronym{dtrdt}{DTrDT}{\gls{dtr} at device level and single-hit
\gls{dt} at block level}
\newacronym{dtrdb}{DTrDB}{\gls{dtr} at device level and single-hit
\gls{db} at block level}
\newacronym{dtrdtr}{DTrDTr}{\gls{dtr} at device level and single-hit
\gls{dtr} at block level}

\newacronym{tc}{TC}{thread cell}
\newacronym{t}{T}{thread}
 \newcommand\kpzOctaBeta{\num{.2414}(2)}

\newcommand\kpzBestAlpha{\num{.3889}(3)}

\newcommand\kpzPPAlpha{\num{.3869}(4)}

\newcommand\kpzAcHRsLam{\num{1.98}(5)}
\newcommand\kpzAcHRsLamZ{\num{1.23}(3)}
\newcommand\kpzAcHRsB{\num{-.4828}(4)}
\newcommand\kpzAcHScaLam{\num{2.01}(2)}
\newcommand\kpzAcHScaLamZ{\num{1.26}(1)}
\newcommand\kpzAcSRsLam{3.8(2)}

\newcommand\kpzAcSRsB{\num{0.76}(2)}

\newcommand\kpzAcSScaLam{\num{1.25}(2)}

 \makeglossaries

\def\figscale{1.2}
\def\figscaledbl{.95}
\def\figscaleqd{.99}

\begin{document}

\title{Dynamical universality classes of simple growth and lattice gas models}

\author{Jeffrey Kelling\textsuperscript{2,3}, G\'eza \'Odor\textsuperscript{1}, and Sibylle Gemming\textsuperscript{3,4}}

\address{\textsuperscript1Institute of Technical Physics and Materials Science,
Centre for Energy Research of the Hungarian Academy of Sciences \\
P.O.Box 49, H-1525 Budapest, Hungary \\}
\address{\textsuperscript2Department of Information Services and Computing, \\
Helmholtz-Zentrum Dresden-Rossendorf \\
P.O.Box 51 01 19, 01314 Dresden, Germany\\}
\address{\textsuperscript3Institute of Ion Beam Physics and Materials Research \\
Helmholtz-Zentrum Dresden-Rossendorf \\
P.O.Box 51 01 19, 01314 Dresden, Germany\\}
\address{\textsuperscript4Institute of Physics, TU Chemnitz\\
09107 Chemnitz, Germany}
\eads{\mailto{j.kelling@hzdr.de}, \mailto{odor@mfa.kfki.hu}}
 
\begin{abstract}
 Large scale, dynamical simulations have been performed for the two dimensional
 octahedron model, describing the \gls{kpz} for nonlinear, or the \gls{ew} class for linear
 surface growth.  The autocorrelation functions of the heights and the dimer
 lattice gas variables are determined with high precision. Parallel \gls{rs} and
 two-sub-lattice stochastic dynamics (SCA) have been compared. The latter causes
 a constant correlation in the long time limit, but after subtracting it one can
 find the same height functions as in case of RS. On the other hand the ordered
 update alters the dynamics of the lattice gas variables, by increasing
 (decreasing) the memory effects for nonlinear (linear) models with respect to
 \gls{rs}.  Additionally, we support the \gls{kpz} ansatz and the Kallabis--Krug
 conjecture in $2+1$ dimensions and
 provide a precise growth exponent value $\beta=\num{0.2414}(2)$. We show the
 emergence of finite size corrections, which occur long before the steady state
 roughness is reached.
\end{abstract}
\maketitle


\glsreset{kpz}
\glsreset{ew}
\glsreset{rs}
\glsreset{sca}
\glsunset{dtrdb}

\section{Introduction}
 Nonequilibrium systems are known to exhibit dynamical scaling,
when the correlation length diverges as $\xi \propto t^{1/z}$,
characterized by the exponent $z$.
Simplest models are \gls{dlg} \cite{Dickmar},
which in certain cases can be mapped onto surface growth
\cite{krug1997review,HalpinHealyZhangReview1995}.
Therefore, understanding \glspl{dlg}, which is far from being trivial
due to the broken time reversal symmetry \cite{Tauber_2014_book},
and is possible mostly by numerical simulations only,
sheds some light on the corresponding interface phenomena \cite{odorbook}.
The simplest example is the \gls{asep} of particles \cite{Spitz70},
in which particles and holes can be mapped onto binary surface slopes
\cite{meakin,PhysRevB.35.3485} and the corresponding continuum model
can be described by the \gls{kpz} equation~\cite{PhysRevLett.56.889}
\begin{equation}  \label{eq:kpz}
\partial_t h(\mathbf{r},t) = \sigma \nabla^2 h(\mathbf{r},t) +
\lambda(\nabla h(\mathbf{r},t))^2 + \eta(\mathbf{r},t) \ ,
\end{equation}
where the scalar field $h(\mathbf{r},t)$ is the height, progressing
in the $D$ dimensional space relative to its mean position, that
moves linearly with time $t$.
This equation was inspired in part by the stochastic Burgers equation
\cite{burgers74} and can describe the dynamics of simple growth processes
in the thermodynamic limit \cite{H90}, randomly stirred fluids~\cite{forster77},
directed polymers in random media \cite{kardar85}, dissipative transport
\cite{beijeren85,janssen86}, and the magnetic flux lines in
superconductors \cite{hwa92}.
In case of surface growth $\sigma$ represents a surface tension,
competing with the  nonlinear up--down anisotropy of strength $\lambda$
and a zero mean valued Gaussian white noise $\eta$. This field exhibits the
covariance
$\langle\eta(\mathbf{r},t)\eta(\mathbf{x^{\prime}},t^{\prime})\rangle =
2 \Gamma \delta^D (\mathbf{r-r^{\prime}})(t-t^{\prime})$.
The $\lambda=0$, linear equation describes the \gls{ew} \cite{EW}
surface growth, an exactly solvable equilibrium system.

Several discrete models obeying these equations have been
studied~\cite{meakin,barabasi,krug1997review}.
The morphology of a surface of linear size $L$ is usually
described by the squared interface width
\begin{equation}
\label{Wdef}
 W^2(L,t) = \frac{1}{L^2} \, \sum_{i,j}^{L} \,h^2_{i,j}(t)  -
 \Bigl(\frac{1}{L} \, \sum_{i,j}^{L} \,h_{i,j}(t) \Bigr)^2 \ .
\end{equation}
In the absence of any characteristic length simple
growth processes are expected to be scale-invariant~\cite{familyVicsek1985}
\begin{equation}
\label{FV-forf}
W(L,t) \propto L^{\alpha} f(t / L^z)\ ,
\end{equation}
with the universal scaling function $f(u)$:
\begin{equation}
\label{FV-fu}
f(u)  \propto
\left\{ \begin{array}{lcl}
     u^{\beta}     & {\rm if} & u \ll 1 \\
     {\rm const.} & {\rm if} & u \gg 1
\end{array}
\right .
\end{equation}
Here $\alpha$ is the roughness exponent in the stationary regime,
when the correlation length $\xi$ has grown to exceeded $L$, and $\beta$ is the
growth exponent,
describing the intermediate time behavior. The dynamical exponent $z$
can be expressed as the ratio of the growth exponents:
\begin{equation}\label{zlaw}
z = \alpha/\beta \
\end{equation}

Apart from the exponents, the shapes of the rescaled width and height
distributions of the interfaces $\Psi_L(\varphi_L)$
were shown to be universal in \gls{kpz} models in both the steady
state~\cite{MPPR02} and the growth regime~\cite{FORWZ1994}.
Here, $\varphi_L$ denotes the interface observable
in question, $W^2$ or $h$, in a system of linear size $L$.
In fact many people define the universality classes by these quantities,
which can be obtained exactly in one
dimension for various surface geometries, like
flat~\cite{CalabreseLeDoussal2011} or
curved~\cite{prahoferSpohn1999,SasamotoSpohn2010,CalabreseLeDoussal2011}
interfaces.
The non-rescaled probability distributions are denoted by $P_L(\varphi_L)$
and their moments are defined via the distribution averages as:
\begin{align}
 \label{eq:distMom}
 \Phi^n_L[\varphi_L] &= \int\limits_0^\infty \left(\varphi_L -\langle
\varphi_L\rangle\right)^n P_L(\varphi_L)\,\mathrm{d}\varphi_L\quad,
\intertext{Two standard measures of the shape, the skewness}
S_L[\varphi_L]&= \langle \Phi^3_L[\varphi_L]\rangle / \langle \Phi^2_L[\varphi_L]\rangle^{3/2}
 \label{eq:skewness}
\intertext{and the kurtosis}
Q_L[\varphi_L]&= \langle \Phi^4_L[\varphi_L]\rangle / \langle
\Phi^2_L[\varphi_L]\rangle^{2} - 3\quad,
 \label{eq:kurtosis}
 \intertext{
are calculated in the steady state or in the growth regime.
The universal, rescaled forms are:
}
\Psi_L[W^2(L)] &=\langle W^2(L)\rangle P_L( W^2(L)/ \langle W^2(L)\rangle)
 \label{eq:kpzPsiWidth}
 \intertext{for the width and}
 \Psi_L[h_L(\mathbf{r})] &= L^\alpha P_L( h_L(\mathbf{r})/ L^\alpha)
 \label{eq:kpzPsiHeight}
\end{align}
for the surface height.
Note, that $\skal{h_L}\equiv \Phi_L^0[h_L]\equiv0$ in the co-moving frame of the
surface.

While many systems are described by a single dynamical length scale,
aging ones are best characterized by two-time quantities, such as
the dynamical correlation and response functions \cite{Cug}.
In the aging regime: $s\gg \tau_{\rm m}$ and $t-s\gg \tau_{\rm m}$,
where $\tau_{\rm m}$ is a microscopic time scale and $s$
is the start time, when the snapshot is taken, one expects the
following law for the autocorrelation function
\begin{align}
 C(t,s) &=
 \skal{\phi(t,\mathbf r)\phi(s,\mathbf r)}-\skal{
 \phi(t,\mathbf r)}\skal{\phi(s,\mathbf r)}\\
&\propto s^{-b} (t/s)^{-\lam_{C}/z}
 \quad,
 \label{eq:kpz_ac_C}
\end{align}
here $\langle\rangle$ denotes averaging over both lattice sites and independent samples;
$\lam_C$ is the autocorrelation and $b$ is the aging exponent.
The function $\phi$ denotes the measured quantity, which can be the particle
density of the lattice gas or the surface height $h(t, \mathbf r)$.
In the latter case,
\begin{align}
 C_h(t,s) &=
 \skal{h(t,\mathbf r)h(s,\mathbf r)}-\skal{h(t,\mathbf r)}\skal{h(s,\mathbf r)}\ ,\\
\intertext{for $t=s$ one finds:}
 &= \skal{h^2(s,\mathbf r)}-\skal{ h(s,\mathbf r)}^2 = W^2(L\to\infty, s) \propto s^{-b_h} \cdot f_C(1)~. \nonumber
\end{align}
This implies the relation
\begin{equation}
 b_h = -2\beta~, \label{eq:kpz_ac_b}
\end{equation}
which must be satisfied in the $L\to\infty$ and $s\to\infty$ limit.
We have also calculated the auto-correlation of the slope
(lattice gas occupancy variables) $n(t,\mathbf{r})$ as:
\begin{align}
C_s(t,s) &=
\left\langle \left( n(t,\mathbf{r} ) - \overline{n} \right)\left( n(s,\mathbf{r}) -
\overline{n} \right)
\right\rangle
\nonumber \\
&= \left\langle n(t,\mathbf{r}) n(s,\mathbf{r}) \right\rangle -
\overline{n}^2
\nonumber \\
&= s^{-b_s} f'_C\left( \frac{t}{s} \right) , \label{Cseq}
\end{align}
where $\overline n = \num{0.5}$ is the conserved average occupancy of sites.
However, $C_s(t,s)$ decays much faster than the height auto-correlator and
obtaining reasonable signal/noise ratio requires much higher statistics.

A dynamic, perturbative renormaliztion group (RG) analysis of the
KPZ equation~\cite{PhysRevE.55.668,PhysRevE.56.1285} suggested that the short
and the long time scaling behavior of the height correlation function are
identical and deduced a scaling relation for the exponent of
$C_h(t\gg s, s\to0) \propto (t/s)^{-\theta}$ as:
\begin{equation} \label{Kre-rel}
\theta = (D + 4) / z - 2 \ .
\end{equation}
Since $\theta = \lambda_C/z + 2\beta$, due to \eqref{eq:kpz_ac_b}
the relation \eqref{Kre-rel} holds exactly in the solvable $1+1$ dimensional
case.
In $D\ge 2$ dimensions perturbative RG can't access the strong coupling
KPZ fixed point \cite{KCW}, thus the validity of this law should be
tested by precise exponent estimates.

A conjecture based on a purely geometric argument, advanced by Kallabis and
Krug~\cite{kallabisKrug1999}, which can also be deduced from the
scaling relation \eqref{Kre-rel} claims $D=\lambda_C$ in any
dimensions. In~\cite{PhysRevE.89.032146} we provided marginal agreement
for this in $D=2$ and Ref.~\cite{carrascoOliviera2014} also suggested
it using solid-on-solid models, although clear power laws were not reached
within the times studied. Here we provide stronger numerical
evidence in case of different lattice gas dynamics.

In aging systems a similar scaling form is expected for the autoresponse
function of the field $\phi$:
\begin{equation}
R(t,s) = \left. \frac{\delta \left\langle \phi(t)\right\rangle}{\delta
j(s)}\right|_{j=0}
 = s^{-1-a} f_R\left(\frac{t}{s}\right)\label{eq:ar} \,
\end{equation}
where $j$ is the external conjugate to $\phi$ and $a$ denotes the so-called
aging exponent $a$. The universal scaling function exhibits the
asymptotic behavior $f_{R}(t/s) \sim (t/s)^{-\lambda_{R}/z}$
with the autoresponse exponent $\lambda_{R}$. In equilibrium
$\lambda_C = \lambda_R$ and $a=b$ due to the fluctuation-dissipation
(FD) symmetry \cite{HP}. In nonequilibrium systems these exponents
can be completely independent. Therefore, we shall determine them
one-by-one and investigate if some extended FD relation may occur
among them. This has been done, using our very recent aging response
exponents \cite{kellingOdorGemming2016_LSI}, determined to test the
validity of a logarithmic extension of \gls{lsi}
\cite{HP} proposed in~\cite{Henkel2013} and work on such extensions in other
models has been continued recently~\cite{Henkel2017_NLMCI,DurangHenkel2017}.

Throughout the present study we compare results obtained using two common
updating schemes for lattice models: \gls{rs} and \gls{sca} (checkerboard)
updates. We find constant as well as non-trivial corrections to the dynamical
correlation functions produced by the \gls{sca} and also observe differences in
the corrections to scaling.

This paper is structured as follows. The investigated model and the
simulation algorithms are introduced in Sect.~\ref{s:models}.
Roughness growth results are presented in Sect.~\ref{ss:resultsR},
while autocorrelation and aging date can be found in Sect.~\ref{ss:resultsAC}.
We conclude the paper with a discussion of the main implications of
our results in Sect.~\ref{s:discussion}.
 
\section{Models and simulation algorithms\label{s:models}}
 Discrete models set up for KPZ have been studied a lot in the
past decades \cite{meakin,barabasi,krug1997review}. A mapping between
KPZ surface growth in two dimensions and driven lattice gases
has been advanced in~\cite{KrugSpohnbook,odor09} an extension of
the "rooftop" model of \cite{meakin,PhysRevB.35.3485}.
We called it octahedron model, characterized by binary slope variables
$\sigma_{x/y}$ at the edges connecting top vertexes of octahedra~\cite{odor09}
representing atoms.
The $\sigma_{x/y}$ take the values $0$ or $1$ to encode down or up slopes,
respectively.
Thus deposition or removal of octahedra corresponds to a
stochastic cellular automaton, with the simple Kawasaki update rules~\cite{odor09}
\begin{equation}\label{rule}
\left(
\begin{array}{cc}
   0 & 1 \\
   0 & 1
\end{array}
\right)
 \overset{p}{\underset{q}{\rightleftharpoons }}
\left(
\begin{array}{cc}
   1 & 0 \\
   1 & 0
\end{array}
\right)\quad,
\end{equation}
where $p$ and $q$ denote the acceptance probabilities.
Projecting the edges onto a plane yields a square lattice of slopes,
which can then be considered as occupancy variables. This
maps the octahedron model onto self-reconstructing
dimers following an oriented migration along the bisection of
the $x$ and $y$ directions of the surface (see figure~\ref{fig:2dscaup} or
figure~\ref{fig:oct} of the supplemental material for a 3D depiction).
In this picture the surface heights must be defined relative to a reference
point $h_{1,1} = 0$ and can be reconstructed from the slope variables
as
\begin{equation}\label{height}
h_{i,j} = \sum_{l=1}^i [2\sigma_x(l,1)-1] + \sum_{k=1}^j [2\sigma_y(i,k)-1] \ .
\end{equation}

Discrete surface and \gls{dlg} models usually apply random sequential dynamics.
On the other hand in certain cases synchronous, so called
\gls{sca}-like site updating can prove to be
useful, especially for simulations on parallel computers. This study
is based on massively parallel simulations on
graphics cards (GPUs). \glsunset{gpu}
Synchronous updating in case of one-dimensional \gls{asep} models has already
been investigated
\cite{aseppardyn,journals/cphysics/SchulzOON11,1742-5468-2012-08-P08004}.
One-point quantities in the bulk, like particle current or
surface growth have been shown to exhibit the same behavior as in
case of \gls{rs}. However, n-point correlation functions may be different.

Here we extend the parallel two-sublattice scheme developed for
\gls{asep} \cite{journals/cphysics/SchulzOON11} to the two dimensional
dimer model as shown on figure~\ref{fig:2dscaup}, and compare the
dynamical scaling results with those of the \gls{rs} dynamics.
\begin{figure}[h]
 \centering
\includegraphics[width=\linewidth/2]{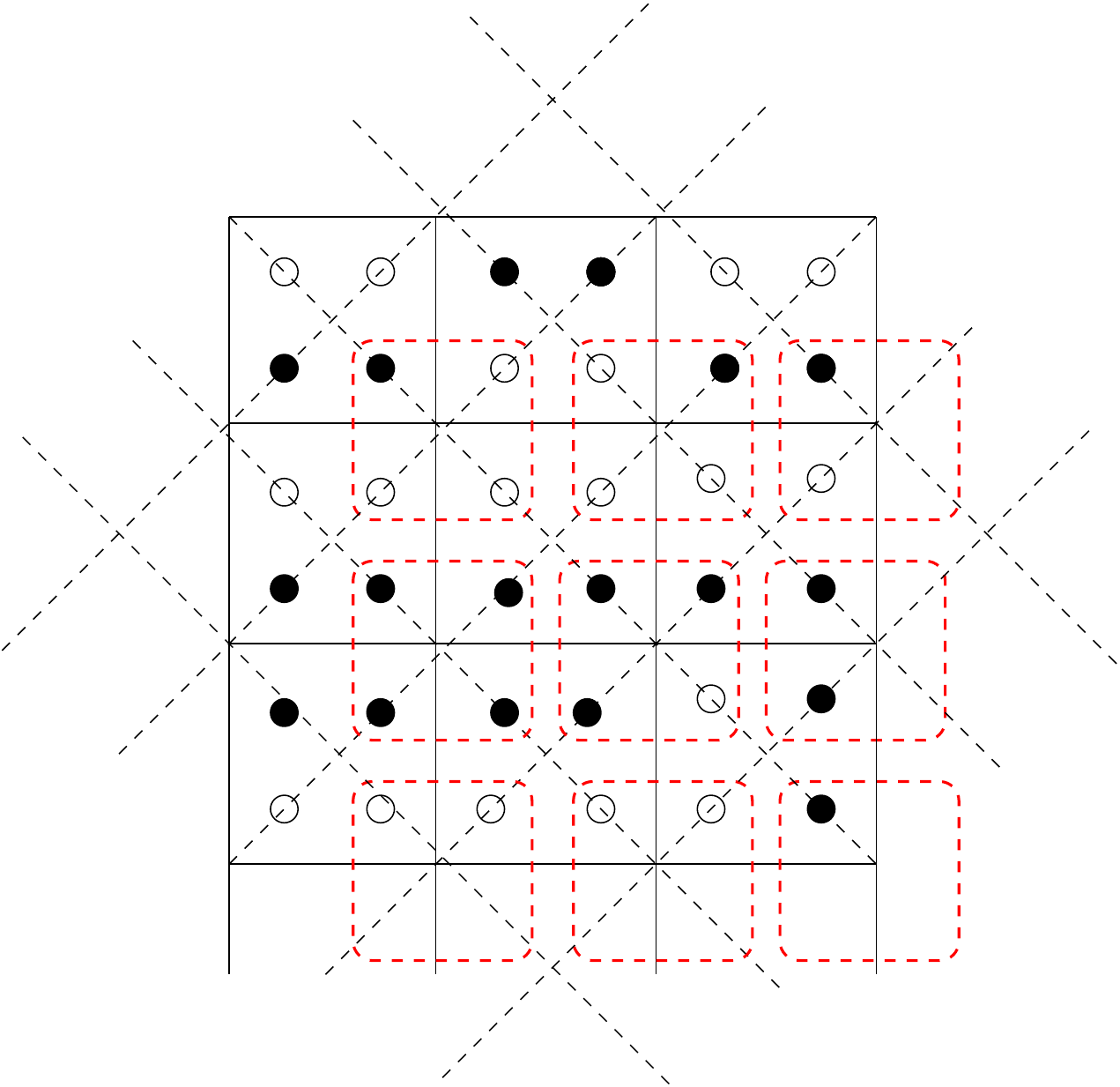}
\caption{Schematics of the two sub-lattice SCA updates of the dimer
lattice gas model. Circles: empty sites (down slopes), bullets: filled
sites (up slopes). Solid, black lines denote areas, where the rule
(\ref{rule}) is applied at $t$ odd, while dashed red lines encircle
areas for update at even $t$ time steps. Diagonal, dashed lines are
parallel with the $x$ and $y$ axis and are projection of the
octahedron edges.}\label{fig:2dscaup}
\end{figure}
While the latter produces uncorrelated deposition and removal processes,
\gls{sca} dynamics attempts updates in a checkerboard pattern, which are thusly
correlated.
Because of this, blocks of sites to be updated can be visited in a sequential
order within a \gls{sca} sub-lattice step, allowing for very
efficient implementation~\cite{MPP2000}, matching
perfectly parallel processors of \gls{gpu}
architectures~\cite{kellingOdorGemming2016_INES}.

Performing \gls{rs} simulations on \glspl{gpu} is less straight-forward, because
unwanted correlations may be introduced~\cite{PhysRevE.89.032146}. In order to
eliminate these and to achieve results as close to really sequential
simulations as possible, we apply a new \gls{dd} scheme, with two layers
of DD to match the \gls{gpu} architecture.
At level one, in a double tiling decomposition, the origin is
moved randomly after each sweep of the lattice (DTr).
At level two, these tiles are subdivided further, with a
logical dead border (DB) scheme. Collective update attempts are
preformed inside these cells, excluding one lattice site-wide
borders around each. Here, the
decomposition origin is moved randomly after each collective update attempt.
This scheme will be referred to as \emph{DTrDB} in the following.
Details of the new implementation are documented
elsewhere~\cite{KellingOdorGemming2017_KPZtech}.

In order to estimate the asymptotic values of different exponents
for $t\to\infty$, local slope analyses of the scaling laws were
performed~\cite{odorbook}. For example in case of the interface
width growth we used
\begin{equation}
\label{eq:beff}
\beta_\mathrm{eff}\left(\frac{t_i - t_{i/2}}2\right) = \frac{\ln W(L\to\infty,t_i) - \ln
W(L\to\infty,t_{i/2})}{\ln(t_i) - \ln(t_{i/2})}\,.
\end{equation}
In our studies the simulation time, measured in \glspl{mcs},
between two measurements was increased exponentially
\begin{equation}
 t_{i+1} = (t_i+10)\mathrm{e}^m\quad,
\end{equation}
using $m=\num{.01}$ and $t_0 = 0$. A flat initial state is realized by a
zig-zag pattern with $W^2(L,t_0)=\num{.25}$. The simulations are subject to
periodic boundary conditions.

Statistical uncertainties
are provided as $1\sigma$--standard errors,
defined as $\Delta_{1\sigma} x = \sqrt{\langle x^2\rangle - \langle
x\rangle^2/(N-1)}$.
Throughout this study we used the implementation of the Levenberg--Marquadt
algorithm~\cite{Levenberg1944,Marquardt1963} in the
gnuplot software~\cite{gnuplot} for non-linear least squares fitting.
  
For $p = q > 0$ the octahedron model describes the surface growth of the
\gls{ew} equation~\cite{odor09} in $2+1$ dimensions.
In this case the autocorrelation function of heights has been
derived~\cite{roethleinBaumannPleimling2006,roethleinBaumannPleimling2006erratum}:
\begin{equation}
 \label{eq:ewAc2d}
 C_\mathrm{h}^{\mathrm{EW}} =  c_0 \ln\kla{\frac{t+s}{t-s}}\quad,
\end{equation}
where $c_0$ is a model-dependent constant. This function approaches $0$ for
$t\gg s$ as a \gls{pl} with the exponent
$\lam_{C,\text{h,\gls{ew}}}/z_\text{\gls{ew}} = 1$, where $z_\text{\gls{ew}}=2$.
In Sect.~\ref{sss:acHeightsEW} we shall reproduce this result numerically as a
test of our simulations.
 
\section{Results\label{s:results}}
 Extensive dynamical simulations were performed using both \gls{rs} and
\gls{sca} updating schemes. To avoid finite-size effects
we considered large systems with lateral sizes of $L=2^{16}$.

In \gls{sca} simulations the deposition probability must be $p<1$
in order to allow stochastic noise. We investigated
three cases: $p=\num{.5}, \num{.75}$ and $\num{.95}$ in depth.
While \gls{kpz} runs were performed without removals: $q=0$,
in the \gls{ew} growth we applied $p=q=1$ for \gls{rs} and
$p=q=\num{.5}$ for \gls{sca}.

The roughness scaling of the interface width is analyzed in
section~\ref{ss:resultsR}. This is followed by autocorrelation and
aging studies of the height as well as lattice-gas variables in
section~\ref{ss:resultsAC}.
  \subsection{Roughness scaling\label{ss:resultsR}}
  \begin{figure}
 \centering
 \includegraphics[scale=\figscaledbl]{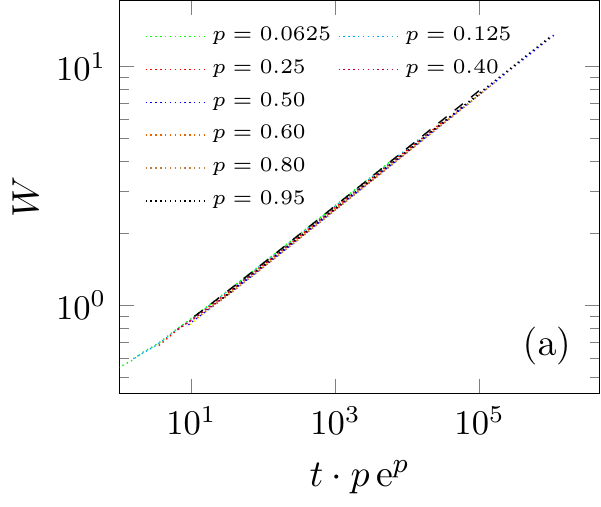}
 \includegraphics[scale=\figscaledbl]{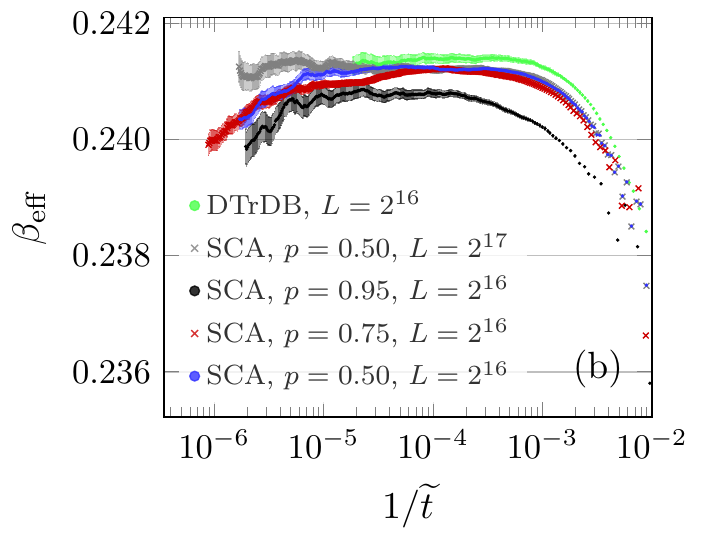}
 \caption[\gls{kpz} state roughness growth under \gls{sca}
 dynamics]{\label{fig:scalingWidthSca}
  Width-scaling under \gls{sca} dynamics. (a) Width curves collapsed over
  $p$ by rescaling time as $\wt t = t\cdot p\,\eu{p}$. For comparison
  the \gls{dtrdb} result is also shown (black dashed line).
  (b) Effective scaling exponents under \gls{sca} dynamics for $p=\num{.95}$
  ($n\geq 2254$),
  $p=\num{.75}$ ($n\geq 6430$) and
  $p=\num{.5}$ ($n\geq 373$, $n\geq 3062$). \gls{rs} data is shown for
  comparison ($n\geq708$).
  Propagated $1\sig$ error bars are attached to the effective exponents, merging
  into an error-corridor at late times due to the dense placing of points.
 }
\end{figure}

To compare numerical results coming from different updates we determined
experimentally a scaling function $f(t,p)$, that provides collapses of
$W(L,f(t,p))$ for different dynamics.
In case or \gls{rs} dynamics this function is linear $f_\mathrm{RS}(t,p) \propto p$.
Since the $p\to 0$ limit of the \gls{sca} corresponds to \gls{rs}
updating, we tried to extend the linear form analytically.
A smaller survey study of \gls{sca} for a larger number of different
$p\leq\num{.95}$ values was used to obtain this function numerically
and resulted in the following nonlinear extension:
\begin{equation}
 f_\mathrm{SCA}(t,p) = \wt t(p) = t\cdot p\,\eu p \ . \label{eq:scaScalingTscale}
\end{equation}
The speedup with respect to linear function of \gls{rs} can be understood
as follows. A dimer, that was moved at a given time step becomes
the target of another update at the next sub-lattice step in the
$p \to 1$, $q=0$ case. This is more effective than random sequential
updating. Therefore, the roughness growth is faster under \gls{sca} than
under \gls{rs} dynamics.
One can test this function by observing a reasonably collapse
on figure~\ref{fig:scalingWidthSca}(a) for different $p$-s of
\gls{sca} as compared to the \gls{rs} results.

Figure~\ref{fig:scalingWidthSca}(b) shows the effective scaling exponents
$\beta_\mathrm{eff}$, as defined in~$\eqref{eq:beff}$, for \gls{sca} and \gls{rs} simulations
as the function of the rescaled time variable.
Most notably, the $\beta_\mathrm{eff}$ exponent results exhibit slightly shifted
plateaus for almost two decades in time, but the difference lies well
within the error-margin of our best published result
$\beta_\mathrm{eff}=\num{.2415}(15)$~\cite{PhysRevE.84.061150}.

\paragraph{\Glsdesc{rs}\label{sss:octaGrowthRegime}}
The pronounced plateau visible for \gls{dtrdb}, suggests that
corrections at these late times are small, thus the $\beta_\mathrm{eff}$
here should be close to the asymptotic value for $\beta$.
This leads to the estimate $\beta=\kpzOctaBeta$, where the
error margin is about the size of the $1\sig$-error bars attached to the
effective exponents at late times.

\paragraph{\Glsdesc{sca}} Like in the \gls{rs} case, there are almost
two decades long plateaus in the effective exponents, depending on
$p$ for $L=2^{16}$. The plateau value differences are beyond the
statistical fluctuations and the \gls{dtrdb} result. The deviation
from the RS result shrinks as we decrease $p$, i.e. as we introduce
more and more randomness. This is plausible, but smaller $p$ also
means less effective simulations.

The plots also show a break down of $\beta_\mathrm{eff}$ at late times. Such
behavior can be attributed to the onset of the steady state, which
is not apparent from existing finite size scaling
studies~\cite{PhysRevE.84.061150}, where $\xi\propto t^z \sim L$ appears to be
reached about one decade later than the left end of the displayed plot.

Most importantly, the $\beta_\mathrm{eff}$ curve does not show this cut-off in
the plateau in case of our largest sized $L=2^{17}$ data, but
matches perfectly the \gls{rs} result. It only shows noise related
oscillations within the $1\sig$-error margin.
This indicates that the cut-off is related to finite sizes that
will be investigated further in the following section.

\subsubsection{Distribution of interface heights in the growth regime}

\begin{figure}[t]
 \centering
 \includegraphics[scale=\figscaledbl]{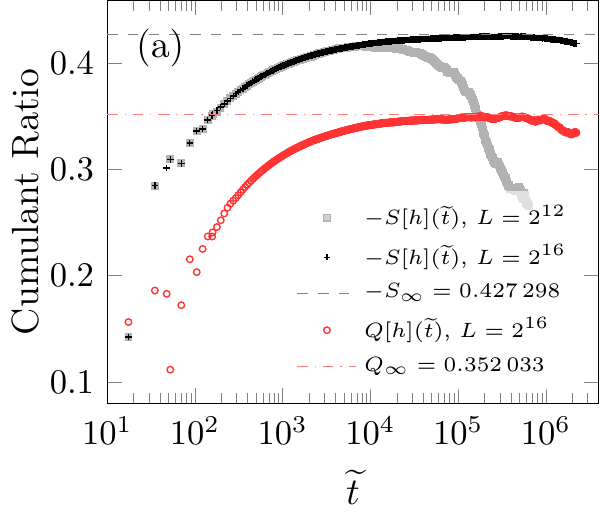}
 \includegraphics[scale=\figscaledbl]{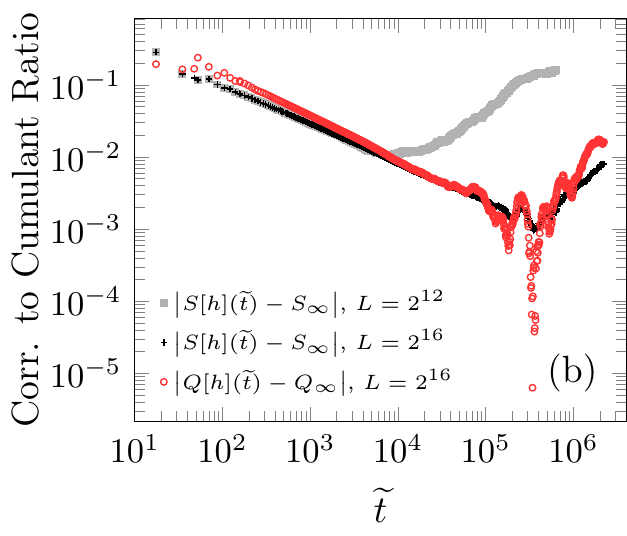}
 \caption[\gls{kpz} growth phase height distribution]{\label{fig:growthHeights}
  Skewness $S[h]$ and kurtosis
  $Q[h]$ of the distribution of interface
  heights in the
  growth regime. The data belongs to the set of \gls{sca} runs with
  $p=\num{.75}$, $L=2^{16}$ ($n_{\mathrm{\gls{sca}, p=\num{.75}}}\geq 6430$, compare
  figure~\ref{fig:scalingWidthSca}). The skewness for
  a smaller dataset for $L=2^{12}$~($n_{\mathrm{\gls{sca}, p=\num{.75},
  L=2^{12}}}\geq 45$), which reaches the steady state regime at late times in
  the plot, is included to illustrate finite-size behavior.
  (a) Cumulant ratios as functions of time. The horizontal lines show the
  obtained fit parameters for the asymptotic values, to guide the eye. See text
  for proper values with error estimates.
  (b) Finite-time and finite-size corrections to the asymptotic
  values of the cumulant ratios.
 }
\end{figure}

In order to get information about the shapes of the distribution of
the \gls{sca} interface heights we calculated their lowest moments for
$L=2^{16}$ and for a smaller $L=2^{12}$, to distinguish finite time from finite
size-corrections.
Figure~\ref{fig:growthHeights} shows the evolution of the cumulant ratios
$S[h]$ and $Q[h]$, defined by~\eqref{eq:skewness} and~\eqref{eq:kurtosis}.
The curves approach their growth regime asymptotic values, but move
away again at late times. These values: $S_\infty$ and
$Q_\infty$ can be determined by performing a fit of the form:
\begin{equation}\label{eq:R}
 R(t) = R_\infty + a_R / \wt t^{2\beta} + b_R / \wt t^{4 \beta}\quad,
\end{equation}
where $\beta$ is the growth exponent, which is motivated by the \gls{kpz} ansatz
discussed in the next section. We use
$\wt t$, defined in~\eqref{eq:scaScalingTscale}, so the timescales match
between \gls{rs} and \gls{sca} run across various deposition probabilities.
In~\eqref{eq:R}, $R$ is a placeholder for $S[h]$ or $Q[h]$, in the interval: $200\leq\wt
t \leq \SI{200000}{MCS}$, which excludes early time oscillations as well as
the cut-off at late times, coming from $\xi\to L$.
This yields $S_{\infty} = \num{-0.427}(2)$ and $Q_{\infty} = \num{.352}(3)$
for the growth regime, in agreement with literature
values~\cite{PhysRevLett.109.170602,Halp13,alvesFerreira2012,AOF2013}
of the \gls{kpz} universality class.
The sign of $S$ depends on the choice $p\gtrless q$ in the simulations,
corresponding to $\sign(\lam)$ of the \gls{kpz} equation~\eqref{eq:kpz}.

Panel (b) of figure~\ref{fig:growthHeights} shows the deviations from these
asymptotic values.
The error estimates given above originate from this representation:
The error is assumed to be on the order of the closest approach
of the numerical data to the asymptotic value.

After the closest approach to the asymptotic values in the growth regime,
$S(\wt t)$ and $Q(\wt t)$, both, move in the direction to their respective values in
the steady state: $S_{\infty}\approx\num{.26}$ and
$Q_{\infty}\approx\num{.13}$~\cite{MPP2000,PaivaReis2007,AaroReis2004,KellingOdorGemming2016_rsos}.
The shape of the distribution of surface heights changing in this way is an indication of
finite-size effects becoming relevant at $\wt t_\mathrm{fs} \approx
\SI{3e5}{MCS}$. This coincides with the time at which the cut-off \footnote{
 The cut-off was observed at $\wt t'\approx\SI{1.7e5}{MCS}$ at two
 different times contributing in the calculation of $\beta_\mathrm{eff}(\wt t')$:
 $\wt t_1\approx \SI{2e4}{MCS}$ and $\wt t_2 \approx\SI{3.2e5}{MCS} \gtrsim
 t_\mathrm{fs}$.
} in $\beta_\mathrm{eff}(\wt t)$ was observed in \gls{sca} runs for $L=2^{16}$
(see figure~\ref{fig:scalingWidthSca}(b)). Hence it becomes clear, that this
change in $\beta_\mathrm{eff}$ is caused by finite-size effects.
In figure~\ref{fig:growthHeights}(b) we also plotted $S[h](\wt t)$ of
a smaller system ($L=2^{12}$), for which the steady state is reached after the
relaxation time $\tau=L^z \approx \SI{6e5}{MCS}$, but finite-size corrections are
evidently relevant long before then.

\subsubsection{KPZ ansatz for the growth regime\label{ss:resultsKpzAnsatz}}

\begin{figure*}
 \centering
  \includegraphics[scale=0.9]{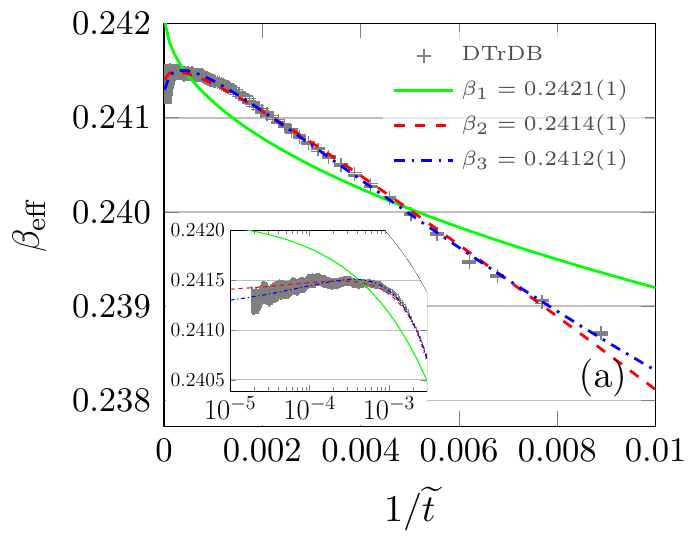}
  \includegraphics[scale=0.9]{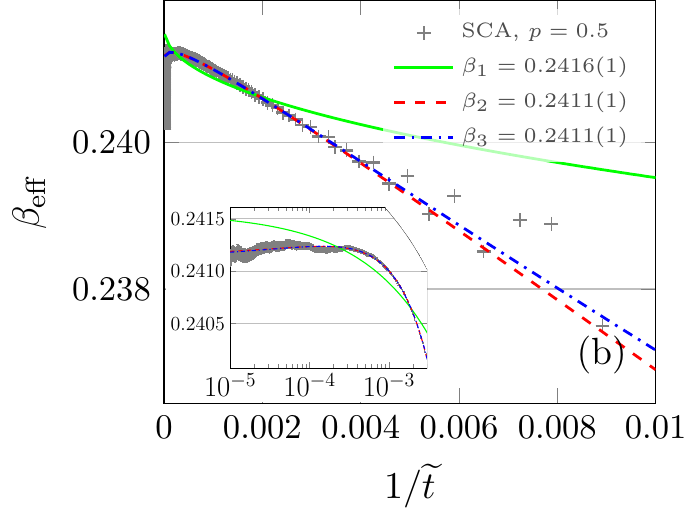}
 \\[2em]
  \includegraphics[scale=0.9]{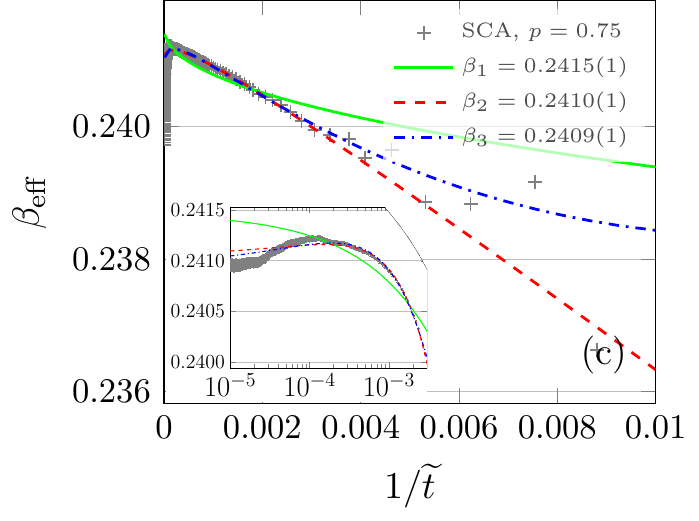}
  \includegraphics[scale=0.9]{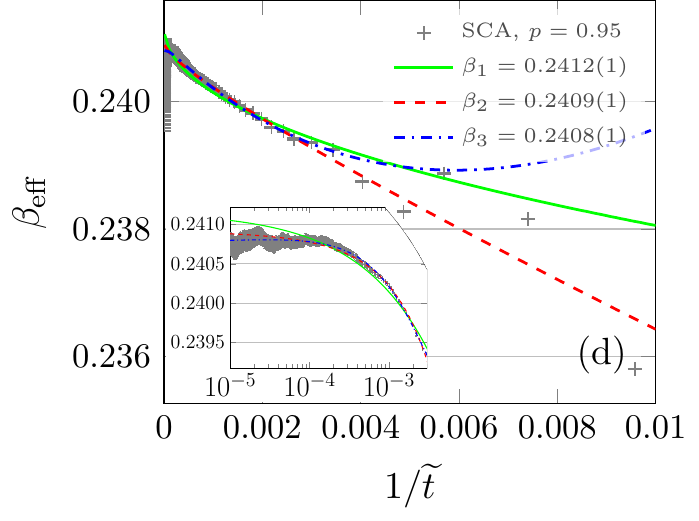}
 \caption[\gls{kpz} ansatz: \gls{rs} vs.~\gls{sca}
 ]{\label{fig:kpzAnsatz}
  Effective exponents $\beta_\mathrm{eff}$ for roughness growth with \gls{kpz}
  ansatz fits using the form~\eqref{eq:betaEffKpzAnsatzEven} to orders one
  through three. The resulting asymptotic values for $\beta$ are given in the
  legends accompanied by the uncertainty of the fit parameter. The insets show a
  zoom to the late-time region $\num{1e-5}\leq1/\wt t\leq\num{3e-3}$ (truncated
  before the finite size break down figure~\ref{fig:scalingWidthSca}(b)).
  Panel (a) shows the \gls{rs} dataset using \gls{dtrdb}. Fits were performed in
  the interval $\num{1e-5}\leq1/\wt t\leq\num{1e-2}$.
  Panels (b)-(d) show
  \gls{sca} datasets with $p=\num{.5}, \num{.75}$ and $\num{.95}$, respectively.
  The fits were restricted to the interval shown in the inset.
  The sample size for \gls{dtrdb}, was $n\geq1044$. See the
  captions of figure~\ref{fig:scalingWidthSca} for other sample sizes.
 }
\end{figure*}

Analytical and numerical investigations of \gls{kpz} models in $1+1$ dimensions
found that finite-time corrections to $h(t)$ took the form $\propto t^{-\beta}$
for the interface
height~\cite{FerrariFrings2011,SasamotoSpohn2010,AOF2014_critDim,AOF2013_nonuniv,HalpinHealyLin2014}:
\begin{align}
 \nonumber
 h(t) &= \sign(\lam)\cdot (\Gamma t)^\beta \chi + \xi + \zeta t^{-\beta}\quad,
 \intertext{where $\lam$, $\Gamma$, $\xi$ and $\zeta$ are model-dependent
  parameters and $\chi$ is a universal random variable with \gls{goe}
  distribution in case of a flat initial condition. The \gls{kpz} ansatz
  hypothesis states, that a generalisation of this form should also hold in
 higher dimensions~\cite{PhysRevLett.109.170602,AOF2013}.
Higher moments of the height $\skal{h^n}$ show corrections $~\propto t^{-n\beta}$,
accordingly, and thus $~\propto t^{-2\beta}$ for the roughness, prescribing:
 }
 \beta_{\mathrm{eff}} &= \beta + \sum\limits_{n=1}^N c_n t^{-2n\beta}\ ,
 \label{eq:betaEffKpzAnsatzEven}
\end{align}
with non-universal parameters $c_n$ and $N$~\footnote{
This assumes that $\xi$ and $\zeta$ are independent, but there is no
guaranty for that. For ballistic deposition in $D=1$, a strange correction
exponent, close to $1/2$ was observed (see e.g.~\cite{AOF2013_nonuniv}),
while in our recent RSOS model simulations~\cite{KellingOdorGemming2016_rsos}
we also found correction exponent $\beta$ in case of $N > 1$ levels.}.
Moreover, good agreement between the numerics and experiments has been found
~\cite{Take10,Take12,carrascoOliviera2014}.
In the $2+1$ dimensional \gls{rsos} model, the dominant corrections to the
roughness growth were found to be of order $~\propto t^{-4\beta}$~\cite{AOF2013},
which motivates the inclusion of higher orders in these forms in higher
dimensions~\cite{PhysRevLett.109.170602,AOF2013,Halp13,halpinhealy2014}.
Ideally, such a model would fit the data well as soon as all relevant orders are
included. Adding more terms should not improve the fit quality further. However,
adding more free parameters in this way can result in overfitting of a
noisy data, if not convergence-problems.

Figure~\ref{fig:kpzAnsatz} shows fitting results using~\eqref{eq:betaEffKpzAnsatzEven}
on the previously introduced datasets.
It is immediately apparent, that \eqref{eq:betaEffKpzAnsatzEven}
with $N=1$ does not describe the presented data, $n=2$-terms are required,
as in case of the \gls{rsos} model.
In case of \gls{rs} simulations, the ansatz appears to fit reasonably well
early times: $\wt t\geq 100$ as well. The \gls{sca} runs on the other hand
show strong oscillations at early times, caused by the synchronous updates,
and are not described well by the \gls{kpz} ansatz here.
Still, late times before the finite size cutoff becomes effective,
(in the interval $\num{1e-5}\leq1/\wt t\leq\num{3e-3}$)
can be fitted well by~\eqref{eq:betaEffKpzAnsatzEven}, suggesting universality
of the corrections. This becomes true in the $p\to 0$ limit, as in case
of figure~\ref{fig:scalingWidthSca}.

The spread of $\beta$ values for larger $N$ provides an estimate for overfitting
and may serve as an error estimate for a small confidence
interval of $1\sigma$. For simulations with \gls{rs} dynamics,
this yields $\beta=\num{0.2414}(2)$, which, remarkably, is identical
to the result
based on the average of the late-time plateaus of $\beta_\mathrm{eff}$.
Fit parameters for $N\leq6$ can be found in table~\ref{tab:kpzAnsatzFits} of the
supplemental material.

\subsubsection{The KPZ dynamical exponent}

The dynamical exponent $z=\alpha/\beta$ of the \gls{kpz} class is related to the
roughness exponent $\alpha$ by the Galilean symmetry~\cite{forster77}:
\begin{equation}
 2 = \alpha + z = \alpha \kla{1 + 1/\beta} \label{eq:galScale} \quad.
\end{equation}
Inserting the our estimate $\beta=\kpzOctaBeta$ into this equation yields
$\alpha=\kpzBestAlpha$ and $z=\num{1.611}(3)$. The latter is used to calculate
autocorrelation exponents in the next section.
It should be noted, that the above value for $\alpha$, while in agreement with
earlier numerical
estimates~\cite{MPP2000,PhysRevE.84.061150,RodriguesOliveira2014,halpinhealy2014},
marginally disagrees with the currently most accepted one
$\alpha=\kpzPPAlpha$~\cite{PaganiParisi2015}. Combining this
roughness exponent results with our own estimate for $\beta$ violates
equation~\eqref{eq:galScale} by about $\num{2.5}\sigma$.
A slight violation of the Galilean invariance, which was proposed for discrete
systems~\cite{WRDE2010}, may explain this disagreement. If this is the
case the correct dynamical exponent would be $z=\num{1.603}(3)$.
However, the validity of the Galilean invariance is still widely accepted in
literature~\cite{PhysRevE.86.051124,RodriguesOliveira2014,barabasi}, for this
reason we use our $\alpha$ and $z$ estimates, obtained using
~\eqref{eq:galScale}, for consistency.
 
\glsunset{tc}
 \subsection{Autocorrelation\label{ss:resultsAC}}
  \subsubsection{Autocorrelation of interface heights in the KPZ
  case\label{sss:acHeightsKPZ}}
   \paragraph{Aging}

\begin{figure}
 \centering
 \includegraphics[scale=\figscale]{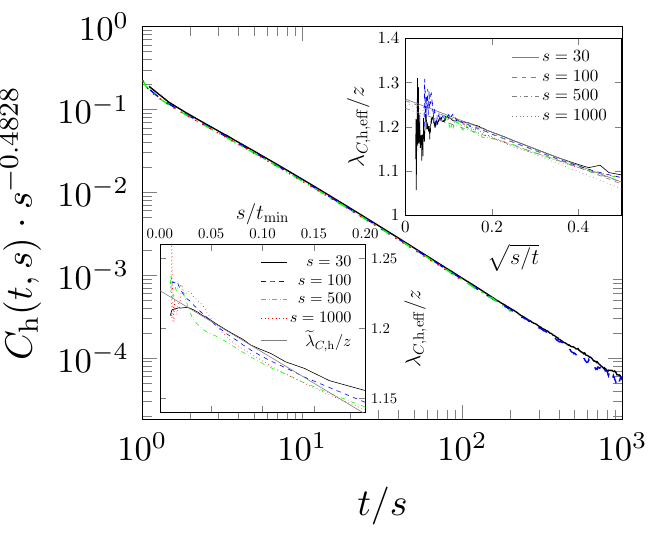}
 \caption[Autocorreclation results for heights from \gls{rs} calculations.]{\label{fig:ach}
  Autocorrelation results from \gls{rs} calculations using \gls{dtrdb}.
  System size $L=2^{16}$, $n\geq1044$ realizations for $s>30$
  and $n\geq473$ for $s=30$.
  The main panel shows the collapsed autocorrelation functions for waiting times
  $s=30,100,500,1000$.
  Right inset: Corresponding local slope analysis and
  extrapolations assuming corrections of the form $\sqrt{s/t}$, as drawn. Linear
  fit was performed for $\sqrt{s/t} \in [\num{.1},\num{.3}]$. Stated errors are
  pure fit-errors, see text for actual error margins.
  Left inset: Tail effective exponents
  obtained from \gls{pl} fits for intervals
  $t \geq t_\mathrm{min}$ with successively increasing $t_\mathrm{min}$. A
  linear fit to the combination of all curves is displayed as a solid black
  line, extrapolating to $\wt\lam_{C,\mathrm{h}}/z = 1.23(3)$.
 }
\end{figure}

The autocorrelation results of the interface heights under \gls{rs}
dynamics are summarized in figure~\ref{fig:ach}.
A near-perfect collapse of the $C_h(t,s)$ functions could be achieved by
using $b$ from the relation~\eqref{eq:kpz_ac_b}.

\paragraph{Autocorrelation exponent obtained by \gls{rs} dynamics}

We calculated effective exponents of $\lambda_{C,h}/z$ and its $t/s\to\infty$
behavior, by an analysis shown in the right inset of figure~\ref{fig:ach}.
In order to read-off the appropriate correction to scaling we linearised
the left tail of the curves by plotting them on the $\sqrt{s/t}$ scale.
This leads to the extrapolations
$\lam_{C,\mathrm{h}}/z = \num{1.254}(9)$ depending on $s$.

To clarify the situation we attempted a different type of local slope
analysis presented in the left inset of figure~\ref{fig:ach}, using tail effective
exponents, where each $\lam_{C,\mathrm{h}}/z$ value was determined
as the exponent of a \gls{pl}-fit to $C_h(t', s)$ for $t' \geq t$.
These can be expected to converge more monotonically to the asymptotic
value as before, because the left tail data of $C_h(t,s)$ are included in
the procedure for all $t_\mathrm{min}$ with an increasing weight
as $t_\mathrm{min}$ increases. Indeed, the curves of different $s$ values
in figure~\ref{fig:ach}(b) behave more linearly with some additional
oscillations. However, all curves seem to fluctuate around a common mean,
which is not the case for the local slope analysis.
A single linear fit for the combination of all curves,
yields an averaged extrapolation of $\wt\lam_{C,\mathrm{h}}/z =
\kpzAcHRsLamZ$ in a marginal agreement with our previous result for
this $\lambda_{C,h}/z = 1.21(1)$ \cite{PhysRevE.89.032146} and
with the value obtained in~\cite{halpinhealy2014,carrascoOliviera2014}
for intermediate times.

The present larger error margin takes the uncertainty
due to the actually unknown corrections into account. This problem is
illustrated in the comparison between effective exponents and tail effective
exponent, where the former show a smaller apparent extrapolation error. In the
following we use the simpler extrapolation method  based on effective exponents,
but estimate the error from their direct fluctuations rather than the
uncertainty of the extrapolation fit.

Using our $z$ value
the corresponding autocorrelation exponent is $\lam_{C,\mathrm{heights}}=\kpzAcHRsLam$.
These results also hold in simulations with more coarse \gls{dd}, where one
would expect an observable difference if any artificial correlations were
present.
    \paragraph{\glsname{sca} autocorrelation functions and aging}

\begin{figure*}
 \centering
  \includegraphics[scale=\figscaleqd]{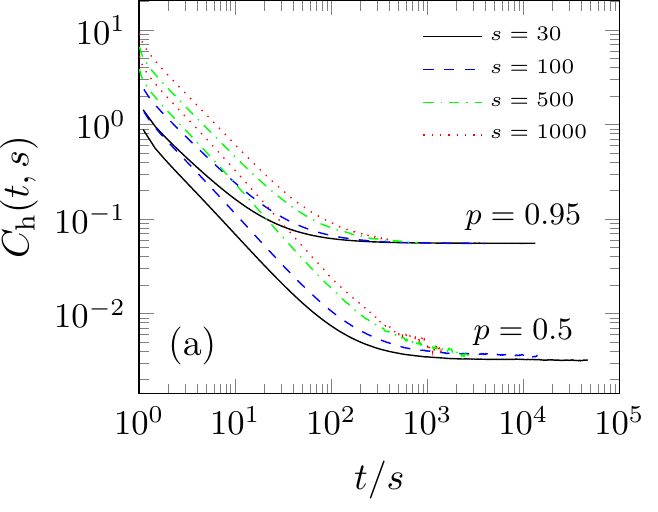}
  \includegraphics[scale=\figscaleqd]{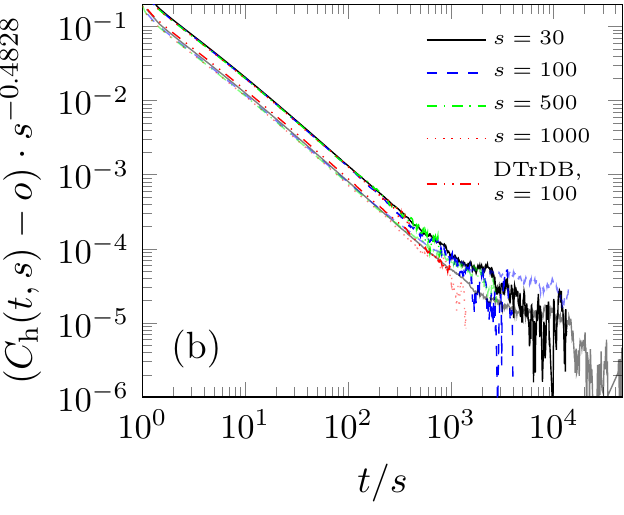}
 \\[2em]
  \includegraphics[scale=\figscaleqd]{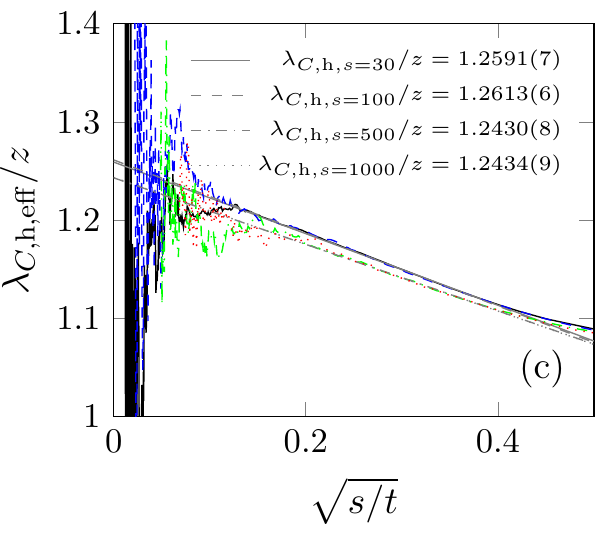}\hspace{1em}
  \includegraphics[scale=\figscaleqd]{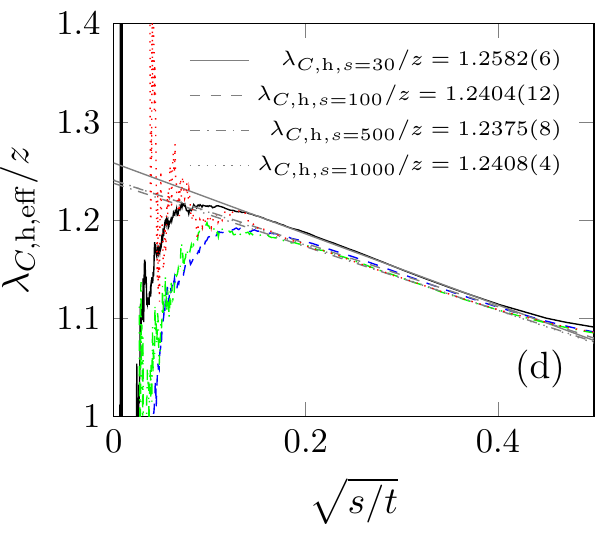}
 \caption[Autocorreclation results for the KPZ heights using \gls{sca}
  updates.]{\label{fig:achSca}
  Autocorrelation of KPZ heights from \gls{sca} calculations. Error
  bars have been omitted for clarity. The visible noise is a good indication
  for $1\sig$ error.
  Panels~(a) and~(b) show
  data sets with $p=\num{.5}$ (3062 realizations, $t\leq\SI{1.4}{MMCS}$) and
  $p=\num{.95}$ (3062 realizations, $t\leq\SI{400}{kMCS}$).
  Lateral system size is $L=2^{16}$.
  (a): Raw autocorrelation functions showing
  saturation depending on $p$.
  (b): Collapsed autocorrelation functions, corrected
  by the saturation offset $o$ (see text). Plots for $p=\num{.5}$ (lower set of
  curves) use paler variations of
  colors than those of $p=\num{.95}$ for the same $s$,
  to make them distinguishable at late times. The \gls{dtrdb}
  autocorrelation function of $s=100$ is also displayed for comparison.
  The bottom panels~(c) and~(d), show the
  local slope analysis corresponding to the $p=\num{.95}$ and $p=\num{.5}$ data
  sets, respectively. Extrapolations assume corrections of the form
  $\sqrt{s/t}$, as drawn. Printed error margins are pure fit-errors.
 }
\end{figure*}

\gls{sca} updates are spatially correlated, therefore they
introduce a contribution to the autocorrelation function, which depends
on the update probability $p<1$. If we want to model cellular automaton
like systems this is not a problem, but for describing the KPZ equation
this is artificial.
Figure~\ref{fig:achSca} compares the autocorrelation functions
of height variables at $p=\num{.95}$ and $p=\num{.5}$.
The most apparent property is the finite asymptotic value
(figure~\ref{fig:achSca}(a)).
This is the consequence of frozen regions, arising in ordered domains,
which are difficult to randomize by the SCA dynamics. In the dimer model
updates can happen at the boundaries only, besides this alternating
domains are also stable in case of SCA, they flip-flop at
even-odd sublattice steps, when $p\to 1$.

We applied an iterative fitting procedure to determine the functional
behavior as follows. As a first approximation the
$C_{\mathrm{h}}(t\to\infty,s,p) = o(p)$ limit was determined using a
linear extrapolation from the function's right tail.
Subtracting the appropriate value from each curve revealed a \gls{pl} approach to
this constant. To obtain refined $o(p)$ values, the exponent $x$
was read off from the data, allowing a subsequent fit for the tail in the form:
\begin{equation}
 f(t) = o + c\cdot t^{-x},
\end{equation}
with free parameters $o$ and $c$. The corrected exponents converged
as $x' \to \lam_C/z$, after subtracting the refined $o(p)$ values.
These iterations yielded self-consistent estimates for $o(p)$ and the
autocorrelation exponent of the \gls{sca}. This procedure is more prone to
statistical error for small $t/s$, because $C_{\mathrm{h}}(t,s)$ is farther away
from the asymptotic behavior in this case, allowing noise in the tail to
influence the extrapolated value more strongly.
Table~\ref{tab:kpzAcScaO} lists the calculated $o(p)$ limits
(including those for the lattice-gas variables, see Section~\ref{sss:scSlopesKPZ}).

The limiting value turned out to depend exponentially on $p$.
Note, that similar $e^p$ dependence has been found in $f_\mathrm{SCA}(t,s)$
relating SCA and RS timescales \footnote{ These limits could also be
determined from the small survey study presented in figure~\ref{fig:scalingWidthSca}(a),
comprising much smaller sample sizes than the results presented in detail in the
following. This data suggests an exponential dependence~$o(p) \propto \exp(\nu p)$
with a similar, or possibly the same, value for the parameter $\nu$ for both
slopes and heights.  However, these autocorrelation measurements used the
same waiting time $s$, without taking into account the $p$-dependent time-scale.
Thus the actual waiting times $\wt s$ decrease with $p$, which makes the fit
performed on the $o(p)$ across these runs unsuitable to determine a reliable
value for $\nu$.
}.
\begin{table}
 \caption[KPZ \gls{sca} autocorrelation limits $o$]{\label{tab:kpzAcScaO}
  Autocorrelation limits for KPZ with \gls{sca} dynamics for different deposition rates
  $p$ and $q=0$, as functions of the waiting time $s$. Fit errors are shown, which
  are below the given number of digits in case of the slopes.
 }
 \tabcapskip
 \centering
 \begin{tabular}{lrrrr}
  \hline
  $s/\SI{}{MCS}$ & \multicolumn{2}{c}{$o_\mathrm{h}$} &
  \multicolumn{2}{c}{$o_\mathrm{s}$} \\
  & $p=\num{.5}$
  & $p=\num{.95}$
  & $p=\num{.5}$
  & $p=\num{.95}$
  \\
  \hline
  30  &\num{0.00320}(3)&\num{0.055398}(8)&\num{0.012871}&\num{0.221623}\\
  100 &\num{0.00331}(5)&\num{0.05520}(3) &\num{0.014286}&\num{0.219827}\\
  500 &\num{0.0031}(2) &\num{0.05457}(8) &\num{0.013944}&\num{0.218547}\\
  1000&\num{0.0035}(3) &\num{0.0548}(2)  &\num{0.013903}&\num{0.218330}\\
  \hline
 \end{tabular}
\end{table}

Figure~\ref{fig:achSca}(b) shows the corrected $C_{\mathrm{h}}(t,s)$ functions,
after subtracting the limiting $o(p)$ values. A nearly perfect data
collapse could be achieved using the aging exponent $b_\mathrm{h}$,
coming from the \gls{rs} simulations.
Even more, the corrected \gls{sca} and the displayed \gls{rs}
autocorrelation functions show identical behavior.

\paragraph{Autocorrelation exponent: \gls{sca}}
Local slope analyses of the corrected autocorrelation functions are displayed in
figures~\ref{fig:achSca}(c) and~\ref{fig:achSca}(d) at $p=\num{.95}$ and
$p=\num{.5}$, respectively. Assuming a rescaling of the abscissa:  $\sqrt{s/t}$,
allows one to observe a linear behavior of the effective exponents for
intermediate times. In case of $p=\num{.95}$ the $o$-values could not be
determined precisely enough for $s>100$, thus we considered extrapolations
at $s=30,100$ only in a weighted average of the results.
This yielded: $\lam_{C,\mathrm{h}}/z=\kpzAcHScaLamZ$
and so $\lam_{C,\mathrm{h}}=\kpzAcHScaLam$.
These values are in good agreement with those obtained from a
local slope analysis of \gls{rs} calculations for small $s$.

The effective exponents for $p=\num{.5}$ show a slightly decreasing
tendency with $s$ in figure~\ref{fig:achSca}(d), moving towards the
\gls{rs} estimate $\wt\lam_{C,\mathrm{h}}/z = \kpzAcHRsLamZ$.
However, we can't consider the extrapolated values for
$s=500$ and $s=1000$ more precise, than those at $s=30,100$, because
the determination of the $o(s)$ constant becomes more uncertain
at higher times, increasing the possible error of the exponent
estimates.
   \subsubsection{Autocorrelation of interface heights in the EW
   case\label{sss:acHeightsEW}}
   \begin{figure}[th!b]
 \centering
 \includegraphics[scale=\figscale]{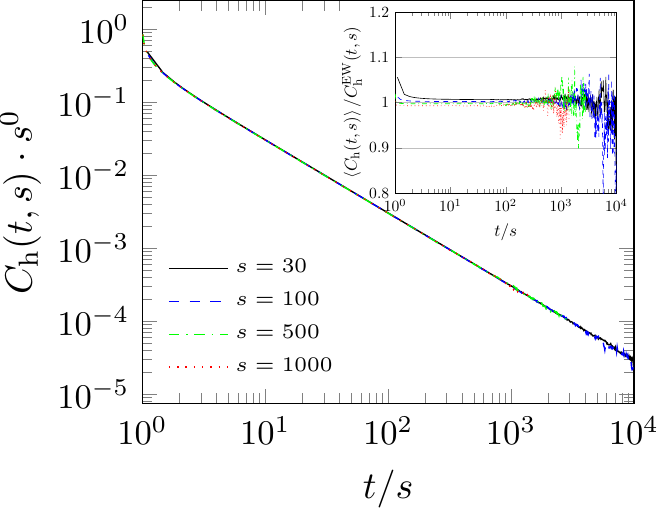}
 \caption[\gls{ew} autocorrelation of heights]{\label{fig:ewAcSca}
  Autocorrelation functions of heights under \gls{sca} dynamics with
  $p=q=\num{.5}$ (EW). Sample size is $n_\text{\gls{sca}}=5919$. Error bars are omitted
  for clarity. The magnitude of fluctuations can be seen from the visible
  fluctuations in the plots.
  The inset shows the data divided by $C_\mathrm{h}^{\mathrm{EW}}$.
 }
\end{figure}

Since the autocorrelation function in the \gls{ew} case is known
exactly~\eqref{eq:ewAc2d}, we can verify our simulations by a
comparison with it. Indeed, the expected form could be reproduced
by our \gls{rs} implementation. A more interesting result is, that
the \gls{sca} simulations also fit it perfectly.
The finite saturation value, caused by correlated updates, observed
in the \gls{kpz} case is not present here.

The agreement with the analytical form is exemplified in the inset of
figure~\ref{fig:ewAcSca}. A small deviation at very early times
can be observed here, as well as in the \gls{rs} results and
should be related to the initial conditions of the simulation
with respect to those of the analytical calculations.
The application of a fit with ~\eqref{eq:ewAc2d} results in
$c_0 \simeq 0.152$ for different waiting times $s$.
Using the consistency relation~(\ref{eq:kpz_ac_b}) for $s\to t$ we can
expect the same value, which was derived for the octahedron model
in~\cite{odor09} for the $s\to\infty$ limit.

These numerical results do not only show the correctness of the
\gls{sca} and \gls{rs} implementations of the roughening kinetics,
but provide an example, where the correlations introduced by \gls{sca}
do not affect the dynamical behavior.

   \subsubsection{Autocorrelation of lattice-gas variables in the KPZ case\label{sss:scSlopesKPZ}}
   \begin{figure}
 \centering
 \includegraphics[scale=\figscale]{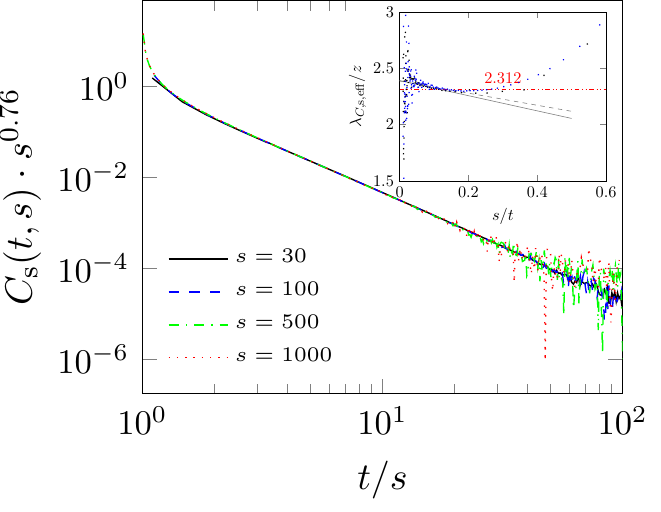}
 \caption[Autocorreclation results for slopes from \gls{rs} calculations.]{\label{fig:acs}
  Autocorrelation results from \gls{rs} calculations using \gls{dtrdb}
  (the same runs as those shown figure~\ref{fig:ach}).
  The main panel shows an aging collapse of the autocorrelation functions for
  $s=30,100,500,1000$.
  The inset displays a local slope analysis for $s=30,100$. Linear fitting lines are also shown,
  assuming corrections of the form ${s/t}$ in the interval $t/s\in[\num{6.25},50]$.
  The horizontal line
  (\tikz[baseline=-\the\dimexpr\fontdimen22\textfont2\relax]{\draw[red,dashdotdotted,](0,0)--(.5,0);})
  marks the value obtained from direct \gls{pl} fits.
 }
\end{figure}

Next we show results for the lattice-gas variables corresponding to the
binary slope values of heights of the KPZ growth presented earlier
(see  figure~\ref{fig:acs}) using \gls{rs} dynamics.
Here again, the $C_s(t,s)$ functions of different waiting times
collapse almost perfectly with the value: $b_\mathrm{s}=\kpzAcSRsB$.
In a previous paper~\cite{PhysRevE.89.032146}
we reported: $b_\mathrm{s}=0.70(1)$, which were obtained by a smaller
sized analysis.

\paragraph{Autocorrelation exponent: \gls{rs}}
Since the density autocorrelation functions decay much more rapidly
than those of the heights, the signal-to-noise ratio in the present
sample is insufficient for a reliable extrapolation based on the
effective exponents. A weighted average of direct
\gls{pl} fits for $4 \leq t/s \leq 90$ yielded
$\lam_{C,\mathrm{s}}/z = \num{2.312}(2)$. However, the effective exponents
show curvature as $t/s\to\infty$ and suggest an asymptotic value
$\lam_{C,\mathrm{s,eff}}/z=\num{2.39}(2)$.
In~\cite{PhysRevE.89.032146} we obtained
$\lam_{C,\mathrm{s,eff}}/z=\num{2.35}(2)$, coming from $s=30$,
$L=2^{13}$ sized CPU simulations.
    \begin{figure*}
 \centering
  \includegraphics[scale=\figscaleqd]{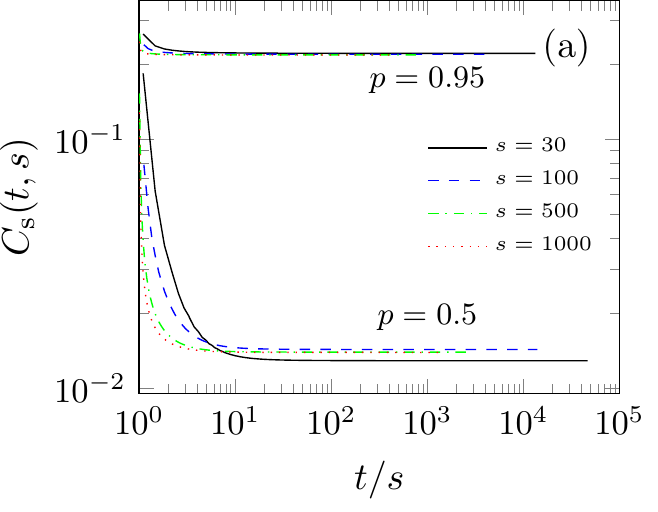}
  \includegraphics[scale=\figscaleqd]{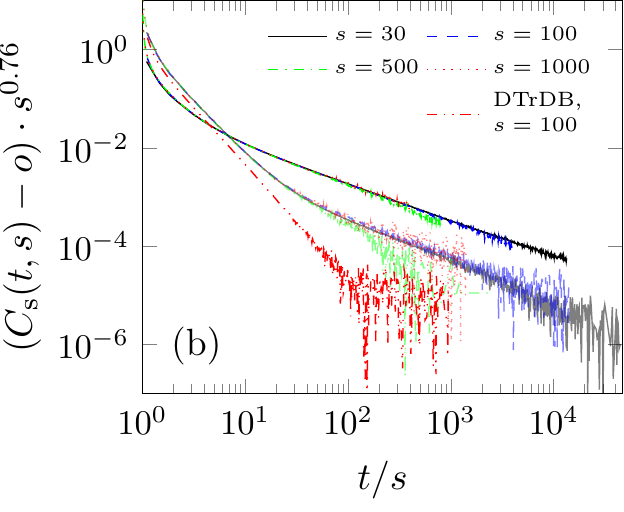}
 \\[2em]
  \includegraphics[scale=\figscaleqd]{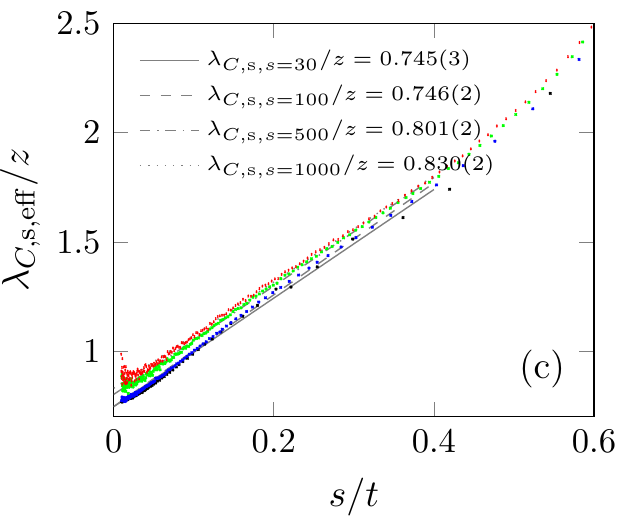}\hspace{1em}
  \includegraphics[scale=\figscaleqd]{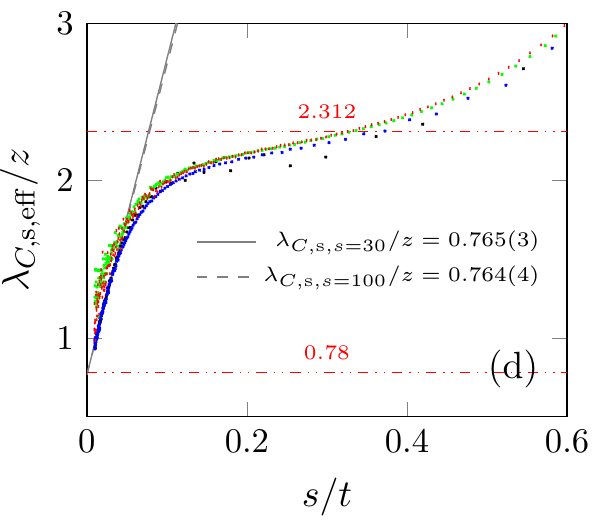}
 \caption[Autocorreclation results for slopes of \gls{sca}
 calculations.]{\label{fig:acsSca}
  Results from \gls{sca} calculations for the autocorrelation of slopes. Error
  bars have been omitted for clarity. The visible noise is a good indication
  for $1\sig$ error.
  Panels~(a) and~(b) show
  data sets with $p=\num{.5}$ and $p=\num{.95}$.
  Data are taken from the same runs as of figure~\ref{fig:achSca}.
  (a): Raw autocorrelation functions showing
  saturation depending on $p$.
  (b): Collapsed autocorrelation functions, corrected
  by the saturation offset $o$ (see text). The lower set (pale curves)
  belongs to
  $p=\num{0.5}$. Data form a \gls{dtrdb} run for $s=100$ is
  displayed in red for comparison.  The bottom
  panels~(c) and~(d),
  show the local slope analysis corrsponding to the $p=\num{.95}$ and
  $p=\num{.5}$ data sets, respectively. Extrapolations assume corrections of the
  form ${s/t}$, as plotted.
  Printed error margins are pure fit-errors.
  Horizontal lines
  (\tikz[baseline=-\the\dimexpr\fontdimen22\textfont2\relax]{\draw[red,dashdotdotted,](0,0)--(.5,0);})
  in panel~(d) mark the asymptotic esponents for
  \gls{rs} updates and \gls{sca} at $p=\num{.95}$, from bottom to top.
 }
\end{figure*}

\paragraph{\glsname{sca} density autocorrelation functions}

Similarly to the case of interface heights the $C_\mathrm{s}(t,s)$ functions
approach finite values asymptotically, as shown in figure~\ref{fig:acsSca}(a)
as the consequence of the \glsname{sca} dynamics.
The computed values of $o(p,s)$ are listed in table~\ref{tab:kpzAcScaO}.

Figure~\ref{fig:acsSca}(b) shows the corrected functions $C_\mathrm{s}(t,s)-o(p)$
in comparison with our \gls{rs} result. Data collapse for $p=0.5$ and $p=0.95$
could be achieved using the common aging exponent value $b_\mathrm{s}=0.76$,
obtained from our previous \gls{rs} calculations.
This indicates, that the density correlation behavior is not changed by the
application
of \gls{sca} updates as in case of the height variables.

\paragraph{\gls{sca} density autocorrelation exponent:}

In contrast with the interface height results the density correlation exponent
of the corrected $C_\mathrm{s}(t,s)$ exhibits a more complex behavior.
The dataset for $p=\num{.95}$ clearly exhibits a different exponent than
what we observed in case of the \gls{rs} simulations.
We show effective exponents fitting for $s=30$ and $100$, where the
signal-to-noise ratio is better on figure~\ref{fig:acsSca}(c).
A direct linear fit extrapolating to $s/t\to0$ yields an estimate of
$\lam_{C,\mathrm{s,\gls{sca}}}/z=\num{.75(2)}$.

At $p=\num{.5}$ we can find a crossover form the \gls{rs} to a different,
\gls{sca} asymptotic behavior in figure~\ref{fig:acsSca}(d).
A linear extrapolation for the tail of this crossover curve results in
$\lam_{C,\mathrm{s,\gls{sca}}}/z$, in good agreement with the
$p=\num{.95}$.
This leads to the following numerical form for the tail of the
autocorrelation function under \gls{sca} dynamics:
\begin{equation}
 f_{C,\mathrm{\gls{sca}}}(t/s,p) \propto c_1 \cdot
 (t/s)^{-\lam_{C,\mathrm{s}}/z} +
  c_2 \cdot (t/s)^{-\lam_{C,\mathrm{s,\gls{sca}}}/z}
  \label{eq:kpzScaAcfC}
\end{equation}
   \subsubsection{Autocorrelation of lattice-gas variables in the EW
   case\label{sss:acSlopesEW}}
   \begin{figure}[tb]
 \centering
 \includegraphics[scale=\figscale]{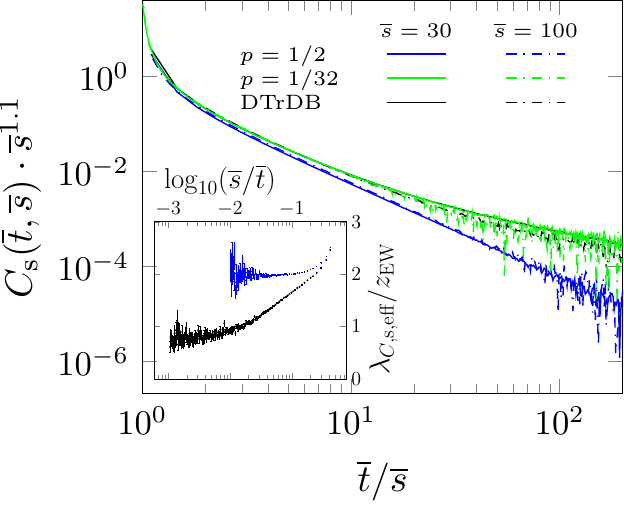}
 \caption[\gls{ew} autocorrelation of slopes]{\label{fig:ewAcS}
  Autocorrelation functions of slopes under \gls{sca} (blue and green) and
  \gls{rs} (black) dynamics in the linear model.
  Sample sizes are $n_{\mathrm{\gls{sca}},p=\num{.5}}=5919$,
  $n_{\mathrm{\gls{sca}},p=1/32}=147$ and $n_{\mathrm{DTrDB}}=2101$, for both
  $\overline s=30$ and $\overline s=100$.
  For \gls{sca}, $p=1/32$, the simulation time is rescaled to collapse the
  curves, following: $\overline t=p\cdot t$, analogously for $s$. No rescaling is
  applied to the other plots: $\overline t=t$.
  Error bars are omitted for clarity. The magnitude of fluctuations can be seen
  from the visible fluctuations in the plots.
  The inset in panel (a) shows the effective autocorrelation exponents
  $\lam_{C,\mathrm{s,eff}}/z_\text{\gls{ew}}$ for \gls{dtrdb} and
  \gls{sca}, $p=1/2$, both for $s=30$ and with $1\sigma$ error
  bars.
 }
\end{figure}

\paragraph{\glsname{rs} autocorrelation functions}
In case of \gls{rs} simulations, the tail of $C_\mathrm{s}(t,s)$ does not decay
with a simple \gls{pl} as can be observed in the figure~\ref{fig:ewAcS}.
The pronounced curvature in the log-log plot suggests a slower than
\gls{pl} decay at first glance. However, the effective exponents (inset)
suggest a \gls{pl} with an asymptotic exponent
$\lam_{C,\mathrm{s}}^\text{\gls{ew}}/z_\text{\gls{ew}}=\num{0.7}(2)$,
following a cross-over from an early-time regime.

\paragraph{\glsname{sca} autocorrelation functions}
While the \gls{sca} dynamics seems to reproduce the expected autocorrelation
function of the surface heights after the removal of the constant,
the evolution of the underlying lattice gas is different.
In case of $p=q=\num{.5}$ the density autocorrelation exhibits a \gls{pl} tail,
characterized by
$\lam_{C,\mathrm{s}}^{\text{\gls{sca}},\num{.5}}/z_\text{\gls{ew}}\approx 2$
(see inset of figure~\ref{fig:ewAcS}).

In the $p\to0$ , $q\to0$ limit the \gls{sca} crosses over to an effective
\gls{rs} dynamics, because we avoid the correlated updates of the lattices.
This is indeed the case here, evidenced by the $p=q=1/32$ results
(see figure~\ref{fig:ewAcS}). Following a rescale of time $t=p\cdot t$ one can
find a good collapse with the \gls{rs} results. Therefore, the update dynamics
seems to affect the scaling behavior of the density autocorrelation function.

\paragraph{Aging}
The aging exponent obtained from the presented simulations is
$b_{\mathrm{s}}^\text{\gls{ew}} = \num{1.1}(2)$. This value holds for both
\gls{rs} and \gls{sca} dynamics, but breaks down for very small values of $s$.
 
\section{Discussion and conclusions\label{s:discussion}}
 \glsresetall
 We performed extensive simulations of the octahedron model by \gls{rs} and
\gls{sca} dynamics. Precise estimates were obtained for the dynamical
behavior: exponents as well as probability distributions of the \gls{kpz}
and \gls{ew} universality classes. The main advance of this work, in the
long story of \gls{kpz} research, is the influence of correlated
\gls{sca} dynamics on the universal properties of these models.
Furthermore, we determined the aging properties of the underlying
\glspl{dlg} of the octahedron model.

By determining moments of the probability distributions we
could study finite size effects and arrived at the conclusion that
the corrections related to this become relevant much before the
occurrence of the steady state.
Our surface growth simulations support the validity of the \gls{kpz} ansatz
hypothesis in $(2+1)D$ and yield a growth exponent $\beta=\kpzOctaBeta$,
from which $\alpha=\kpzBestAlpha$ and $z=\num{1.611}(3)$ can be deduced.
The growth exponent value lies within the error margins of
\cite{KellingOdorGemming2016_rsos,Halp13,PhysRevE.84.061150}, but not within
those of the early landmark result of Forrest and Tang
$\beta=\num{.240}(1)$~\cite{FT90}. However, the small simulation cells used then
demanded shorter simulations times, which could have lead to a smaller estimate
due finite-time corrections.
Under \gls{sca} dynamics marginally lower
growth exponents were observed for deposition probabilities $p>\num{.5}$ and
additional corrections to scaling caused the failure of the \gls{kpz} ansatz
at early times.

Our estimate of the roughness exponent
does not agree with the direct estimate $\alpha=\num{.3869}(4)$, obtained
recently through a finite-size scaling analysis of the \gls{rsos} model
by Pagnani and Parisi~\cite{PaganiParisi2015}, which was based on \gls{sca}
simulations with $p=\num{0.5}$. Numerical differences between \gls{sca} and
\gls{rs} dynamics might be a cause of this. However, since our estimate was derived
using~\eqref{eq:galScale}, a
slight violation of the Galilean invariance, which was proposed for discrete
systems~\cite{WRDE2010}, may also explain this disagreement.

Both our \gls{rs} and \gls{sca} simulations reproduced the expected
autocorrelation behavior of interface heights in the \gls{ew} universality class.
In the \gls{kpz} case correlated updates resulted in $C_\mathrm{h}(t,s)$ to approach a
finite value asymptotically. However, after the subtraction of this constant
we found the same universal \gls{pl} tails for both types of site-selection dynamics.

In case of the underlying lattice-gas variables, we found the relevance of the
\gls{sca} dynamics for the asymptotic autocorrelation decay exponents,
but the aging exponent seems to be insensitive for this. Interestingly,
in case of the non-linear (\gls{kpz}) model the \gls{sca} dynamics slows the decay
of the autocorrelations, while in the linear (EW) model this results in a
shorter memory of the dimer model. This is the consequence of the effectivity
of the ordered \gls{sca} updates, which enhances the build up (\gls{kpz})
or distortion (\gls{ew}) of homogeneous areas, correlated for long times.

Our estimates for the autocorrelation exponents of the \gls{kpz} class are
summarized in table~\ref{tab:kpzAC}. We provided numerical results for $C(t,s)$
in the \gls{kpz} case with unprecedented accuracy, drawn from timescales
up to $t/s=1000$ due to the high signal-to-noise ratios we could achieve
by these parallel algorithms implemented on GPUs. These simulations
can be help to test predictions of theories like local scale-invariance
with logarithmic corrections~\cite{Henkel2013}.
  \begin{table}
 \caption{\label{tab:kpzAC}
  Summary of \gls{kpz} and \gls{ew} autocorrelation $\lam_C$ and aging $b$
  exponents, assuming $z_\mathrm{KPZ} = \num{1.611}(3)$ and
  \mbox{$z_\mathrm{EW}=2$},
  respectively.
  There are no independent estimates for $b$ under \gls{sca} dynamics. Values
  for the \gls{ew} case provided without error margin correspond to the
  analytical
  solution~\cite{roethleinBaumannPleimling2006,roethleinBaumannPleimling2006erratum}.
 }
 \tabcapskip
 \centering
 \begin{tabular}{llrrrr}
  \hline
  && $\lam_{C,\mathrm{h}}$ &
  $\lam_{C,\mathrm{s}}$
  & $b_\mathrm{h}$ & $b_\mathrm{s}$
  \\
  \hline
  \\[-2.2ex]
  \multirow{2}{*}{\rotatebox{90}{KPZ}}&
  \gls{rs}
  & \kpzAcHRsLam & \kpzAcSRsLam
  & \kpzAcHRsB& \kpzAcSRsB
  \\
  &\gls{sca}
  & \kpzAcHScaLam & \kpzAcSScaLam
  \\
  \hline
  \\[-2.2ex]
  \multirow{2}{*}{\rotatebox{90}{EW}}&
  \gls{rs}
  & 2 & $\num{1.4}(4)$
  & \multirow{2}{*}{0} & \multirow{2}{*}{1.1(2)}
  \\
  &\gls{sca}
  & 2 & $\approx\num{4}$
  \\
  \hline
 \end{tabular}
\end{table}

The KPZ autocorrelation exponent in $(1+1)$ dimensions was derived analytically
$\lam^\mathrm{1d}_{C,\mathrm{h}} = 1$~\cite{PhysRevE.55.668,PhysRevE.56.1285}. Later Kallabis and
Krug conjectured, that in higher dimensions
$\lam_{C,\mathrm{h}} = D$~\cite{kallabisKrug1999} applies, but rigourous
proof is still missing. Our estimates for $\lam_{C,\mathrm{h}}$,
summarized in table~\ref{tab:kpzAC}, support this hypothesis within
error margin both for \gls{rs} and \gls{sca} dynamics.

We have tested the validity of the relation~\eqref{Kre-rel} by
Krech with our numerical data. The value for the short time dynamical exponent:
$\theta = \lambda_C/z + 2\beta = 1.23(2) + 0.2414(2) \simeq 1.71(3)$
agrees well with: $(D + 4) / z - 2 = 6 / 1.611(3) - 2 \simeq 1.72(1)$,
therefore we can support the validity of the relation obtained by
a perturbative RG analysis~\cite{PhysRevE.55.668,PhysRevE.56.1285}.

A possible continuation of this work could be the study of the
height correlations in the momentum space:
\[
 C_q(t,s) = \left\langle h_{-q}(t) h_q(s)\right\rangle \quad.
\]
In particular one should be able to test, whether $C_q(t,s)$ decays
in an exponential or in a stretched exponential way as predicted in references
\cite{schwartzEdwards2002,EdwardsSchwartz2002,ColaioriMoore2001,katzavSchwarz2004}.
Note, that we have already succesfully used the extension of
the dimer model to determine the power spectrum density:
$S(k,t) = \left\langle h_{-q}(t) h_q(t)\right\rangle$
in case of Kuramoto-Sivashinsky type of systems \cite{OdorLiedkeHeinig2010}.

We can also compare the present estimates with our recently published values for the
autoresponse $\lam_R=\num{2.00}(6)$ and the corresponding aging exponent
$a=0.24(2)$~\cite{kellingOdorGemming2016_LSI}. $\lam_C \simeq \lam_R$ seems
to hold within error margins.
In $(1+1)$ dimensions an exceptional \gls{fdr} exists~\cite{dekerHaake1975,forster77}:
\begin{equation}
 T \chi(t,s;r) = -\partial^2_r C(t,s;r)
 \label{eq:kpz1dFdr} \quad .
\end{equation}
This implies the exponent relations $\lam_C = \lam_R$ and
\begin{equation}
 1 + a = b + 2/z \label{eq:kpz1dAging}
\end{equation}
confirmed by simulations~\cite{PhysRevE.85.030102}.
Our $(2+1)D$ results support the first one, but the latter is not
satisfied by our numerics:
\begin{align*}
 1 + a &= 2 \kla{ \beta + \beta/\alpha} \\
 \num{1.24}(2) &\neq \num{1.724}(3)
\end{align*}
This calls for the existence of a possible \gls{fdr} in higher
dimensions. For example the genaralized form
\begin{equation}
 1 + a + (D-1)/2 = b + 2/z \label{eq:kpz1dAging}
\end{equation}
is satisfied by the exponents within error limits in $D=1,2$ both.
Confirmation of this assumption should be a target of further research.
An intermediate step in this direction could also occur as an inequality, like
one found in the \gls{kpz} steady
state~\cite{KatzavSchwartz2011DynIneq,KatzavSchwartz2011ExpIneq}.

The autocorrelation and aging exponents which we found for the driven lattice
gas of slopes differ from another two dimensional extension of the \gls{tasep}
described in~\cite{PhysRevE.83.051107}, where $\lambda_{C,\mathrm{s}}/z = 1$ and
$b_\mathrm{s}=1$ are reported.

Finally we point out that the \gls{sca} simulations are more efficient because
they they allow for optimal memory access patters in contrast to the random
accesses required for the \gls{rs} ones. Technical details of our implementations
are published elsewhere~\cite{kellingOdorGemming2016_INES,KellingOdorGemming2017_KPZtech}.
The extension of these algorithms for other surface models,
like those with conservation laws \cite{OdorLiedkeHeinig2010,applied} or
in higher dimensions \cite{PhysRevE.81.031112} is straightforward. However, the
efficiency of \gls{rs} implementations, using the approach employed here, decreases
with the number of dimensions due to the volume of local cells increasing.
\gls{sca} simulations do not suffer from this problem and are thus
more suitable for higher dimensional problems.

The code used in this work can be found at~\texttt{https://github.com/jkelling/CudaKpz}.
 
\section*{Acknowledgments}

We are grateful for the useful comments from Malte Henkel and Timothy
Halpin-Healy and thank Herbert Spohn, Giorgio Parisi and Uwe Täuber for helpful
discussions.
Support from the Hungarian research fund OTKA (Grant No.~K109577), the
Initiative and Networking Fund of the Helmholtz Association via the W2/W3
Programme \mbox{(W2/W3-026)} and the International Helmholtz Research School
NanoNet \mbox{(VH-KO-606)} is acknowledged.
We gratefully acknowledge computational
resources provided by the HZDR computing center, NIIF Hungary and the Center for
Information Services and High Performance Computing (ZIH) at TU Dresden.

 \bibliography{article}
 \bibliographystyle{iopart-num}

\clearpage
\title[Supplemental Information]{Supplemental Information:\\Dynamical universality classes of simple growth and lattice gas models}
\appendix
\renewcommand\thesection{Supplement \Alph{section}}

\section{Octahedron Model}
\begin{figure}[hb]
 \centering
 \includegraphics{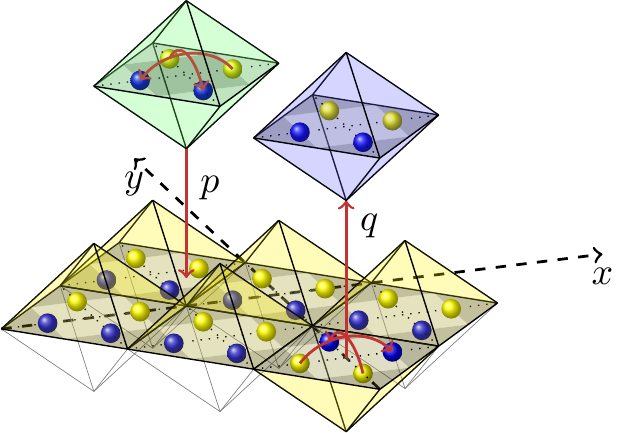}
 \caption{\label{fig:oct}
  Illustration of the lattice gas (blue and yellow balls) in the
  $2+1$--dimensional octahedron model. The depiction maps the particles to
  the up- (blue) and down- (yellow) edges of octahedra. The exchanges of
  slope-pairs connected to deposition ($p$) and removal
  ($q$) processes are illustrated.
  The checkerboard decomposition followed by \gls{sca} updates in indicated in
  the $x$-$y$-plane.
 }
\end{figure}

Figure~\ref{fig:oct} shows a depiction of the update process in the octahedron model
after~\cite{odor09}.

\section{KPZ ansatz for the growth regime}

In the main manuscript we test how the \gls{kpz}
ansatz~\cite{FerrariFrings2011,SasamotoSpohn2010,AOF2013_nonuniv,AOF2014_critDim,
PhysRevLett.109.170602,Halp13,halpinhealy2014}
hypothesis describes the late-time
corrections to the growth exponent $\beta$. Its most general form gives
\begin{align}
 \beta_{\mathrm{eff},h} &= \beta + \sum\limits_{n=1}^N c_n t^{-n\beta}\quad,
 \label{eq:betaEffKpzAnsatzGeneral}
 \intertext{which for $N=1$ is postulated to describe the corrections to the
 growth law for the average surface height.
 Only the special form for the roughness growth, which is reduced to
 exponents which are even multiples of $\beta$, was considered in the main text:}
 \beta_{\mathrm{eff}} &= \beta + \sum\limits_{n=1}^N c_n t^{-2n\beta}
 \label{eq:betaEffKpzAnsatzEven}\quad.
\end{align}

\glsreset{kpz}
\glsreset{ew}
\glsreset{rs}
\glsreset{sca}
\glsunset{dtrdb}
\begin{table*}[bt]
 \caption[\gls{kpz} ansatz: fit parameters for the octahedron model]
 {\label{tab:kpzAnsatzFits}
  Summary of fitting of $\beta_\mathrm{eff}$ using the \gls{kpz} ansatz
  in the most general form~\eqref{eq:betaEffKpzAnsatzGeneral}
  and using~\eqref{eq:betaEffKpzAnsatzEven} with even powers,
  more suitable to describe corrections of the roughness-scaling up to
  maximum orders $N=6$.
  Error margins given are uncertainties of the fit parameters, not actual
  error estimates for $\beta$. The reduced sums of residuals $\chi_\mathrm{red}$
  are given to judge the quality of the fits. For each dataset we underlined
  the value of $\chi_\mathrm{red}$, where the quality does not improve
  significantly by increasing $N$.
 }
 \centering
 \renewcommand{\arraystretch}{1.2}
 \begin{tabular}{lrrrrrrrrrr}
  \hline
  &\multicolumn{4}{c}{\gls{rs}}
  &\multicolumn{6}{c}{\gls{sca}}\\
  &\multicolumn{2}{c}{\gls{dtrdb}, \texttt{TC=1,1}}
  &\multicolumn{2}{c}{\gls{dtrdb}, \texttt{TC=2,2}}
  &\multicolumn{2}{c}{$p=\num{.5}$}
  &\multicolumn{2}{c}{$p=\num{.75}$}
  &\multicolumn{2}{c}{$p=\num{.95}$}\\
  \hline
  $N$&
  $\beta$ & $\chi_\mathrm{red}$ &
  $\beta$ & $\chi_\mathrm{red}$ &
  $\beta$ & $\chi_\mathrm{red}$ &
  $\beta$ & $\chi_\mathrm{red}$ &
  $\beta$ & $\chi_\mathrm{red}$ \\
  \hline
  \multicolumn{11}{c}{equation~\eqref{eq:betaEffKpzAnsatzGeneral}}\\
$1$	&$\num{0.2430}(\num{1})$	&\num{10.04}
	&$\num{0.2433}(\num{1})$	&\num{9.24}
	&$\num{0.2420}(\num{1})$	&\num{7.03}
	&$\num{0.2419}(\num{1})$	&\num{8.63}
	&$\num{0.2418}(\num{1})$	&\num{3.15}
\\
$2$	&$\num{0.2396}(\num{1})$	&\num{1.82}
  &$\num{0.2396}(\num{1})$	&\underline{\num{1.29}}
	&$\num{0.2403}(\num{1})$	&\num{1.62}
	&$\num{0.2400}(\num{1})$	&\num{1.31}
  &$\num{0.2403}(\num{1})$	&\underline{\num{0.51}}
\\
$3$	&$\num{0.2411}(\num{1})$	&\underline{\num{0.79}}
	&$\num{0.2408}(\num{1})$	&\num{0.55}
  &$\num{0.2414}(\num{1})$	&\underline{\num{0.64}}
  &$\num{0.2404}(\num{1})$	&\underline{\num{1.18}}
	&$\num{0.2401}(\num{1})$	&\num{0.49}
\\
$4$	&$\num{0.2421}(\num{2})$	&\num{0.67}
	&$\num{0.2421}(\num{1})$	&\num{0.32}
  &$\num{0.2414}(\num{1})$	&\underline{\num{0.64}}
	&$\num{0.2385}(\num{1})$	&\num{0.63}
	&$\num{0.2403}(\num{2})$	&\num{0.48}
\\
$5$	&$\num{0.2386}(\num{3})$	&\num{0.47}
	&$\num{0.2409}(\num{2})$	&\num{0.28}
	&$\num{0.2388}(\num{3})$	&\num{0.51}
	&$\num{0.2394}(\num{3})$	&\num{0.61}
	&$\num{0.2399}(\num{5})$	&\num{0.48}
\\
$6$	&$\num{0.2377}(\num{8})$	&\num{0.47}
	&$\num{0.2403}(\num{5})$	&\num{0.28}
	&$\num{0.2333}(\num{9})$	&\num{0.45}
	&$\num{0.2420}(\num{7})$	&\num{0.59}
	&$\num{0.2485}(\num{8})$	&\num{0.40}
\\
\hline
\multicolumn{11}{c}{equation~\eqref{eq:betaEffKpzAnsatzEven}}\\
$1$	&$\num{0.2421}(\num{1})$	&\num{7.28}
	&$\num{0.2422}(\num{1})$	&\num{6.27}
	&$\num{0.2416}(\num{1})$	&\num{5.22}
	&$\num{0.2415}(\num{1})$	&\num{6.54}
	&$\num{0.2412}(\num{1})$	&\num{1.76}
\\
$2$	&$\num{0.2414}(\num{1})$	&\underline{\num{1.16}}
 &$\num{0.2414}(\num{1})$	&\underline{\num{1.10}}
 &$\num{0.2411}(\num{1})$	&\underline{\num{0.65}}
	&$\num{0.2410}(\num{1})$	&\num{1.72}
 &$\num{0.2409}(\num{1})$	&\underline{\num{0.81}}
\\
$3$	&$\num{0.2412}(\num{1})$	&\num{0.65}
	&$\num{0.2412}(\num{1})$	&\num{0.33}
	&$\num{0.2411}(\num{1})$	&\num{0.64}
 &$\num{0.2409}(\num{1})$	&\num{1.51}
	&$\num{0.2408}(\num{1})$	&\num{0.52}
\\
$4$	&$\num{0.2413}(\num{1})$	&\num{0.53}
	&$\num{0.2412}(\num{1})$	&\num{0.29}
	&$\num{0.2412}(\num{1})$	&\num{0.59}
 &$\num{0.2407}(\num{1})$	&\underline{\num{0.86}}
	&$\num{0.2407}(\num{1})$	&\num{0.50}
\\
$5$	&$\num{0.2412}(\num{1})$	&\num{0.51}
	&$\num{0.2412}(\num{1})$	&\num{0.28}
	&$\num{0.2412}(\num{1})$	&\num{0.59}
	&$\num{0.2406}(\num{1})$	&\num{0.68}
	&$\num{0.2405}(\num{1})$	&\num{0.41}
\\
$6$	&$\num{0.2412}(\num{1})$	&\num{0.51}
	&$\num{0.2412}(\num{1})$	&\num{0.28}
	&$\num{0.2411}(\num{1})$	&\num{0.57}
	&$\num{0.2403}(\num{1})$	&\num{0.50}
	&$\num{0.2407}(\num{1})$	&\num{0.38}
\\
  \hline
 \end{tabular}
\end{table*}
 
A list of best fit parameters for various versions with $N\leq6$ is provided in
table~\ref{tab:kpzAnsatzFits}, which also lists the reduced sums of residuals
$\chi_\mathrm{red}$ to quantify agreement between the fitting model and the
data.

Where the \gls{kpz} ansatz does indeed apply, fits of the more general
models~\eqref{eq:betaEffKpzAnsatzGeneral} should not show increased agreement
with the data. The table shows them to be less consistent with respect to the
resulting estimates for $\beta$. They provide the best description of the data
with only the term $\propto\wt t^{-2\beta}$, but one or two additional \emph{odd}
terms present ($N=2,3$). The best fits resulting from
models~\eqref{eq:betaEffKpzAnsatzEven} are consistently better than those
of~\eqref{eq:betaEffKpzAnsatzGeneral}, across all datasets, which justifies
discarding the latter class of models and thereby supports the \gls{kpz} ansatz
hypothesis for the roughness growth.

\section{Distribution of interface heights in the growth regime}

In the main manuscript, the finite time corrections to the cumulant ratio of the
distribution of surface heights are shown to be described by the \gls{kpz}
ansatz, yielding the form:
\begin{align}\label{eq:R}
 R_\mathrm{KPZ}(t) &= R_\infty + a_R \cdot \wt t^{\beta} + b_R \cdot \wt t^{2 \beta}\quad.\\
 \intertext{Assuming a single \gls{pl} in the form:}
 R(t) &= R_\infty + a_R \cdot \wt t^{c_R}\quad,
\end{align}
a fit can also be made, resulting in the exponents $c_S \approx\num{-.54}$ and
$c_Q \approx\num{-.60}$ for $S[h](t)$ and $Q[h](t)$, respectively. The value
$c_R\approx \num{-.6}$ also holds for a number of other dimensionless cumulant
rations, not displayed here. The values for $S_\infty$ and $Q_\infty$ resulting
from this extrapolation stay well within the margins of error given in the main
manuscript.
 
\end{document}